\documentclass{aa} 
\usepackage{amsmath}
\usepackage{pdfpages}
\usepackage{subfig}
\usepackage{graphicx}
\usepackage{txfonts}
\usepackage{natbib}
\bibpunct{(}{)}{;}{a}{}{,}

\begin{document}
   \title{Transiting exoplanets and magnetic spots characterized with optical interferometry}

   \subtitle{}

   \author{R.Ligi\inst{1}, D. Mourard\inst{1}, A.-M. Lagrange\inst{2}, K. Perraut\inst{2}, A. Chiavassa\inst{1}}

   	\institute{Laboratoire Lagrange, UMR 7293 UNS-CNRS-OCA, Boulevard de l'Observatoire, CS 34229, 06304 NICE Cedex 4, France.\\
    \email{roxanne.ligi@oca.eu}
  	\and  UJF-Grenoble1/CNRS-INSU, Institut de Plan\'etologie et d'Astrophysique de Grenoble, UMR 5274, Grenoble, F-38041, France }

   \date{Received 16 April 2014 / Accepted 17 October 2014}


  \abstract
   {Stellar activity causes difficulties in the characterization of transiting exoplanets. In particular, the magnetic spots present on most exoplanet host stars can lead to false detections with radial velocity, photometry, or astrometry techniques. Studies have been performed to quantify their impact on infrared interferometry, but no such studies have been performed in the visible domain. This wavelength domain, however, allows reaching better angular resolution than in the infrared and is also the wavelength most often used for spectroscopic and photometric measurements.}
   {We use a standard case to completely analyse the impact of an exoplanet and a spot on interferometric observables and relate it to current instrument capabilities, taking into account realistic achievable precisions.}
   {We built a numerical code called \texttt{COMETS} using analytical formulae to perform a simple comparison of exoplanet and spot signals. We explored instrumental specificities needed to detect them, such as the required baseline length, the accuracy, and signal-to-noise ratio. We also discuss the impact of exoplanet and spot parameters on squared visibility and phase: exoplanet diameter and size, exoplanet position, spot temperature, star diameter.}
   {According to our study, the main improvement to achieve is the instrument sensitivity. The accuracy on squared visibilities has to be improved by a factor 10 to detect an exoplanet of $0.10$~mas, leading to $<0.5\%$ precision, along with phase measurements of $\sim 5^{\circ}$ accuracy beyond the first null of visibility. For an exoplanet of $0.05$~mas, accuracies of $\sim 0.1\%$ and $\sim 1^{\circ}$ from the first null are required on squared visibilities and phases. Magnetic spots can mimic these signals, leading to false exoplanet characterization. Phase measurements from the third lobe are needed to distinguish between the spot and the exoplanet if they have the same radius.}
	{By increasing interferometer sensitivity, more objects will become common between interferometric and photometric targets. Furthermore, new missions such as PLATO, CHEOPS, or TESS will provide bright exoplanet host stars. Measurements will thus overlap and provide a better characterization of stellar activity and exoplanet.}

\authorrunning{Ligi et al.}
   \maketitle
%

\section{Introduction}

The discovery of the first exoplanet around a solar-type star by \cite{Mayor1995} has opened up a new field of research in both planetary and stellar sciences. The numerous exoplanets found up to now show a wide variety in size, composition, and distance from their host star that naturally raises the question of habitability along with the necessity of characterizing these new worlds. We have thus moved from the era of exoplanet discovery to the era of exoplanet characterization, which is closely related to the characterization of host stars.

When we observe an exoplanetary system, the measurements include both the exoplanetary signal and the stellar signal, which depends on several parameters such as activity, rotation period, and inclination. We thus need to interpret these measurements and relate each signal to its physical origin to correctly detect or characterize the exoplanet signal.

Interferometry is of great interest in the direct characterization of exoplanets at many levels. It is a useful tool for excluding false-positive scenarii, which is a recurring problem because unresolved companions mimic a planetary transit for \textit{Kepler} candidates, as shown by \cite{Huber2012}, for example. These authors performed interferometric observations with PAVO \citep{Ireland2008} on CHARA to confirm the hypothesis of a planetary companion around Kepler$-21$. This exoplanet, discovered by \cite{Howell2012}, is an Earth-like planet, and such exoplanets generally provoke smaller variations in radial velocity (RV) measurements than stellar activity, which limits their detection. Beyond the validation of exoplanetary systems, interferometry has long been known as a possible way to detect light from exoplanets \citep{Zhao2008, Monnier2004}, but this still remains difficult. In fact, some attempts to directly detect exoplanets with infrared interferometry have already been made. \cite{Matter2010} on AMBER/VLTI \citep{Petrov2003} and MIDI/VLTI \citep{Leinert2003}, and \cite{Zhao2011} on MIRC/CHARA both tried to directly detect a planetary companion using differential phase and closure phase measurements. Still, instrumental effects limit their detection (for both, a factor of 6 to 10 in the precision of the phase measurements has to be reached to detect the planetary signal), but improvements in their stability should provide better results in the future. 

One of the difficulties encountered in achieving an exoplanet characterization is to measure complex visibilities at advantageous spatial frequencies. This problem has been discussed by \cite{Chelli2009} when they presented the phase-closure nulling method. They showed that the effect of a faint companion is stronger around the visibility nulls of the primary. The companion's contribution remains constant to the phases as long as it is not resolved, including the frequencies of the primary nulls, and is even stronger than the systematic errors. \cite{Duvert2010} applied this method to detect the 5-mag fainter companion of HD$59717$, and derived the mass and size of this binary system.

Finally, interferometry allows measuring many observables of a transiting planetary system through closure-phase measurements. In particular, it allows extracting the inclination angle of the exoplanet \citep{vanBelle2008, Zhao2008} and other important parameters: the impact parameter, the transit velocity, the stellar and planetary radius, and the transit ingress time \citep{vanBelle2008}, which makes this technique a fundamental tool when no other data of the system are available.
 
In some cases, stellar spots can help determining exoplanet properties such as their obliquity \citep[][]{Nutzman2011,Sanchis2011}. In turn, exoplanet studies sometimes allow detecting and characterizing a host star's spots and rotation \citep{Silva2011}. However, most of the time, stellar activity adds complexity to the analysis of the signals. Its characterization is thus necessary but not yet achievable, as shown in \cite{Baron2012} in the context of the Imaging Beauty Contest $2012$. 

The emergence of hypertelescopes contitutes an encouraging option to resolve this problem. It promises the direct imaging of transiting exoplanets and stellar surfaces \citep{Labeyrie2012a, Labeyrie2012b} thanks to kilometric baselines and an important gain in the limiting magnitude of the insturments. \citet{Baines2007} demonstrated the interest of measuring the angular diameter of the star in transiting systems. We thus decided to focus our study on visible interferometric measurements of transiting systems, which has great advantages both scientifically and technically. Since a stellar photosphere is not perfectly symmetrical and we do not necessarily know the position of the transiting exoplanet at the time of the measurement, the information can only be extracted by simultaneously investigating many data. This implies measuring many visibilities and phases and using a \textit{(u,v)} plane that covers many directions. 
The advantage of the visible domain lies first in the increased angular resolution for a given baseline, and second, the fact that it is the main domain used in photometry or in RV for indirect searches for planets. It is already used by several instruments that reach unprecedent angular resolution (up to $0.3$ milliarsecond): VEGA/CHARA \citep{Mourard2009, Ligi2013}, PAVO/CHARA \citep{Ireland2008} and VISION/NPOI \citep{NPOI}. With the significant improvement in sensitivity and accuracy that visible interferometry has gained recently, it might be a good way to characterize spot and exoplanet signatures on interferometric observables and to separate the different signals. 

In Sect.~\ref{sec:Code}, we describe a new numerical code called \texttt{COMETS} that allows modeling interferometric observables of a star hosting a magnetic spot and/or a transiting exoplanet. We apply this code to analyze the impacts that exoplanet and spot parameters produce on the shortest baselines length required to measure their signal in Sect.~\ref{sec:GeneralStudy} and investigate their detectability in Sect.~\ref{sec:Detectability}. We discuss whether it might be possible to distinguish between spot and exoplanet in Sect.~\ref{sec:Detectionwithactivity}. We end by discussing the current interferometer capabilities of detecting exoplanets in Sect.~\ref{sec:discussion}.


\section{Models of the objects}
\label{sec:Code}

\texttt{COMETS} (code for modeling exoplanets and spots) is a numerical code using analytical formulae developed to model the interferometric observables (visibilities, closure phases, or phases) provided by a star with a (dark) magnetic spot and/or a transiting exoplanet. The computation is made on a grid of a \textit{(u, v)} plane with $u$ and $v$ ranging from $-1 km/\lambda$ to $1 km/\lambda$ with a step $du=dv$ smaller than $4 m/\lambda$ (refined if necessary). We use a wavelength $\lambda$ of $720$~nm, as used by VEGA instrument.

\subsection{Representation of a transiting exoplanet}
\label{subsec:exoplanet}

The star is represented by a linear limb-darkened (LD) disk of angular diameter $\theta_{\star}$ obtained with the Claret coefficient $b$ \citep{Claret2011} and maximum radius $\sigma_{\rm max} = \theta_{\star} /2$. Its surface brightness distribution depends on its effective temperature $T_{\rm eff, \star}$ and on the considered point at the stellar surface $\sigma$.

We consider the star centered on the origin of the coordinate system in the sky $\sigma_0 = (\alpha_0,\delta_0 )$. The stellar intensity $I_{\star}$ at any location point depends on an impact factor $\mu$ defined as $\mu = \sqrt{1-\left(\frac{\sigma}{\sigma_{\rm max}}\right)^{2}}$. It represents the cosine of the angle formed by the line of sight and the normal to the surface at the considered point. Thus, the stellar profile intensity can be written $I_{\star}(\mu) = I_\star(1) - b(1-\mu)$ in a linear representation, where $I_\star(1)$ is the stellar intensity at the center of the star.

The contrast between a transiting exoplanet and its host star is very high in the optical domain \citep[typically, the contrast is $10^{10}$ in the visible and $10^{6.5}$ in the infrared for Earth-like planets, see, e.g.,][]{Traub2002}. The transiting exoplanet located at any position $\sigma_{\rm p} = (\alpha_{\rm p},\delta_{\rm p})$ on the stellar disk is thus assumed to be a dark disk of intensity $I_{\rm p}=0$ and angular diameter $\theta_{\rm p}$.
Since the star is a LD disk, the star's intensity at the planet location $I_{\star}(\mu_{\rm p})$ precisely depends on the planet's location. Thus, the intensity profile of the system $star+planet$ can be written:

\begin{equation}
 I_{\star}(\mu) \Pi_{(0,0,\theta)} - \left( I_{\star}(\mu_{\rm p}) - I_{\rm p} \right) \Pi_{(\alpha_{\rm p},\delta_{\rm p},\theta_{\rm p})}\ ,
\end{equation}
where $\Pi$ is the Gate function.

The complex visibility function represents the Fourier transform of the surface brightness distribution of the source. We calculate it in Appendix~\ref{annexPlanet}. Since interferometers mainly measure the squared modulus of object visibilities, we considered the squared modulus of the complex visibility in our study and its argument, called the phase $\phi$, described as $\phi = \tan^{-1}\left(\frac{\Im(V)}{\Re(V)} \right)$, where $\Re$ and $\Im$  define the real and imaginary part of the complex number $V$. For a symmetric uniform disk representation, the amplitude of $V$ follows the Bessel function and decreases in the first lobe before reaching zero at $\theta B / \lambda = 1.22$ (called first null), where the phase jumps from $0$ to $\pm \pi$. From the 2D complex visibility maps generated by \texttt{COMETS}, it is possible to estimate closure phases through calculating the closure equation on any arbitrary triplet of ($u,v$) coordinates. This method has recently been used by \cite{Chiavassa2014}. However, in the present study we decided to limit the analysis to the phase plane without calculating the closure phase. This simplifies the calculation and is a good compromise with respect to the expected effects on the phase.

\subsection{Representation of a spot}
\label{subsec:Spot}

The spot is usually decomposed into two distinct parts, the umbra and the penumbra, of intensities $I_{\rm um}$ and $I_{\rm pen}$. 

A spot located at the position $\sigma_{\rm s} = (\alpha_{\rm s},\delta_{\rm s})$ on the stellar disk is represented by two superimposed disks both centered on $\sigma_{\rm s}$, with $\theta_{\rm pen} > \theta_{\rm um}$, $\theta_{\rm um}$ and $\theta_{\rm pen}$ being the angular diameters of the umbra and the penumbra.

Thus, the intensity profile of the umbra is written $\left( I_{\star}(\mu_{\rm s}) - I_{\rm um} \right) \Pi_{(\alpha_{\rm s},\delta_{\rm s},\theta_{\rm um})}$, $\mu_{\rm s}$ representing the impact parameter at the spot location. The intensity profile of the penumbra is written as the subtraction of two disks of intensity $I_{\rm pen}$ and angular diameters $\theta_{\rm pen}$ and $\theta_{\rm um}$: $ (I_{\star}(\mu_{\rm s}) - I_{\rm pen}) \left( \Pi_{(\alpha_{\rm s},\delta_{\rm s},\theta_{\rm pen})} - \Pi_{(\alpha_{\rm s},\delta_{\rm s},\theta_{\rm um})} \right)$.
The final intensity profile of the system $star+spot$ is thus:
\begin{equation}
\begin{aligned}
& I_{\star}(\mu) \Pi_{(0,0,\theta_{\star})} \\
& - \left( I_{\star}(\mu_{\rm s}) - I_{\rm pen} \right) \left( \Pi_{(\alpha_{\rm s},\delta_{\rm s},\theta_{\rm pen})} - \Pi_{(\alpha_{\rm s},\delta_{\rm s},\theta_{\rm um})} \right) \\
& - \left( I_{\star}(\mu_{\rm s}) - I_{\rm um} \right) \Pi_{(\alpha_{\rm s},\delta_{\rm s},\theta_{\rm um})}
 \ .
\end{aligned}
\label{eq:IntensiteTache}
\end{equation}

The corresponding complex visibility calculation is presented in Appendix~\ref{annexSpot}.
Interferometers cannot see details such as the penumbra, and we therefore ignore it for the rest of this study. Its intensity is set to  $I_{\rm pen}=I_{\star}$ and its diameter to $\theta_{\rm pen}=0$.
To calculate the spot intensity $I_{\rm s} = I_{\rm um}$, we consider it as a black body. We thus calculate the spectral radiance of a spot with Planck's law, which depends on the spot temperature at the chosen wavelength. We normalize it with the stellar radiance (also considered as a black body) to derive the spot intensity such as $0<I_{\rm s}<1$ as needed in the complex visibility equation.

\subsection{Model of a transiting exoplanet and a spot}

The last case to be studied is a star with a transiting exoplanet and a spot. We make the hypothesis that the exoplanet and the spot projected surfaces on the stellar surface do not overlap, that is $N' \neq N$ with $N = \cos(2 \pi (u \alpha_{\rm p} + v \delta_{\rm p})) + i \sin (2 \pi (u \alpha_{\rm p} + v \delta_{\rm p}))$ and $N' = \cos(2 \pi (u \alpha_{\rm s} + v \delta_{\rm s})) + i \sin (2 \pi (u \alpha_{\rm s} + v \delta_{\rm s}))$, and that both are located inside the stellar disk.

From merging the two previous models, the intensity profile of the system $star+exoplanet+spot$ is:
\begin{equation}
\begin{aligned}
&I_{\star}(\mu) \Pi_{(0,0,\theta)} - \left( I_{\star}(\mu_{\rm p}) - I_{\rm p} \right) \Pi_{(\alpha_{\rm p},\delta_{\rm p},\theta_{\rm p})} \\
& - \left( I_{\star}(\mu_{\rm s}) - I_{\rm pen} \right) \left( \Pi_{(\alpha_{\rm s},\delta_{\rm s},\theta_{\rm pen})} - \Pi_{(\alpha_{\rm s},\delta_{\rm s},\theta_{\rm um})} \right) \\
& + \left( I_{\star}(\mu_{\rm s}) - I_{\rm um} \right) \Pi_{(\alpha_{\rm s},\delta_{\rm s},\theta_{\rm um})}
 \ .
\end{aligned}
\label{equ:intensityProfileTotal}
\end{equation}

The corresponding complex visibility calculation is presented in Appendix~\ref{annexPlanetSpot}.

\section{Detecting a transiting planet or a spot}
\label{sec:GeneralStudy}

\begin{figure}[h!]
\includegraphics[scale=0.5]{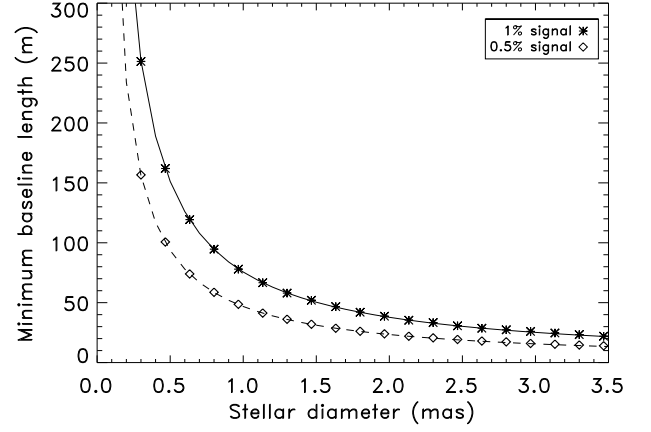} 
\includegraphics[scale=0.5]{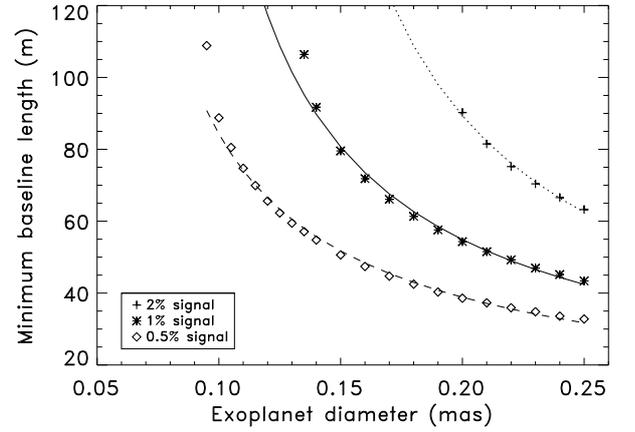} 
\includegraphics[scale=0.5]{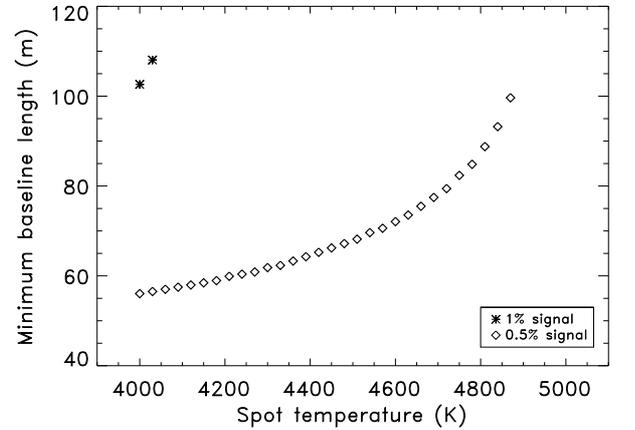}
\caption{Squared visibilities. The MBL is calculated for variations of $2 \%$, $1 \%$, and $0.5 \%$ (crosses, asterisks, and diamonds, respectively) as a function of three parameters. \textit{Top}: stellar diameter with $\theta_{\rm p}/\theta_{\star}=0.15$ and ($\alpha_{\rm p}=0.2\theta_{\star}, \delta_{\rm p}=0$)~mas. \textit{Middle}: exoplanet diameter with $\theta_{\star}=1$~mas and ($\alpha_{\rm p}=0.2, \delta_{\rm p}=0.0$)~mas. \textit{Bottom}: spot temperature for $\theta_{\rm s}=0.15$~mas and ($\alpha_{\rm s}=0.2, \delta_{\rm s}=0.0$)~mas. The curves in the upper panel correspond to the analytical formula (see text). The curves in the middle panel correspond to the fit of Eq.~\ref{equ:EquationGenerale}. }
\label{fig:MBLSqrdVis}
\end{figure}

\subsection{Impacting parameters}

We first performed simulations to estimate the minimum baseline length (MBL) needed to obtain a given signal on the interferometric observables. We have explored the parameter space that can affect the MBL, either for phases or squared visibilities:  exoplanet diameter, stellar diameter, spot temperature, and exoplanet location. However, all parameters but one have to be fixed to study its specific impact on the MBL. Thus, the nominal values of the parameters, when their impact on the MBL is not tested, are given in Table~\ref{tab:FixedParameters}. The stellar diameter ($\theta_{\star}=1$~mas) was chosen to be easily resolved in the visible wavelength with baselines of $\sim 300$m and roughly corresponds to F, G, K stars around which exoplanets have been found until now \cite[see, e.g.,][]{Ligi2012}. The exoplanet and spot location on the stellar disk are the same: ($\alpha_{\rm p,s}=0.2, \delta_{\rm p,s}=0$). The exoplanet diameter corresponds to the upper limit of existing exoplanet angular diameters (see Sect.~\ref{sec:discussion}). 

\begin{table}[h]
\centering
\begin{tabular}{l c}
\hline \hline
Parameter & Fixed value \\
\hline
$\theta_{\star}$ & $1$~mas \\
$T_{\star}$ & $5500$ K \\
$\theta_{\rm p}$ & $0.15$~mas \\
$\theta_{\rm s}$ & $0.15$~mas \\
$T_{\rm s}$ & $4300 K$ \\
\hline 
\end{tabular}
\caption{Values of the parameters when their impact on the MBL is not tested.}
\label{tab:FixedParameters}
\end{table}

\begin{table}
\centering
\begin{tabular}{l c}
\hline \hline
Parameter & Variation domain \\
\hline
$\theta_{\rm p}$ & $0.04-0.24$~mas \\
$\alpha_{\rm p}$ & $0-0.30$~mas \\
$\theta_{\star}$ & $0.30-3.35$~mas \\
$T_{\rm s}$ & $4000-5170$ K \\
\hline 
\end{tabular}
\caption{Variation domain of the parameters when their impact on the MBL is tested.}
\label{tab:ParametersVariation}
\end{table}

We have chosen realistic variation domains for each parameter. They are given in Table~\ref{tab:ParametersVariation}. The upper limit of the exoplanet diameter lies beyond the generally known exoplanet angular diameters, and would correspond more to a brown dwarf, but a signal corresponding to such a large object in interferometric observables would allow distinguishing between a planetary and a stellar companion. When testing the impact of the stellar diameter, the ratio $\theta_{\rm p}/\theta_{\star}$ is steady and equal to $0.15$, which would be the situation if the same system were seen at different distances, otherwise two parameters would vary at the same time (the stellar diameter and that of the exoplanet). The exoplanet location is proportional to the stellar diameter, that is ($\alpha_{\rm p}=0.2 \times \theta_{\star}, \delta_{\rm p}=0.0$). The spot's temperature variation is set according to \cite{Strassmeier2009} and \cite{Berdyugina2005}.

\begin{figure*}
\begin{center}$
\begin{array}{cc}
	\includegraphics[scale=0.5]{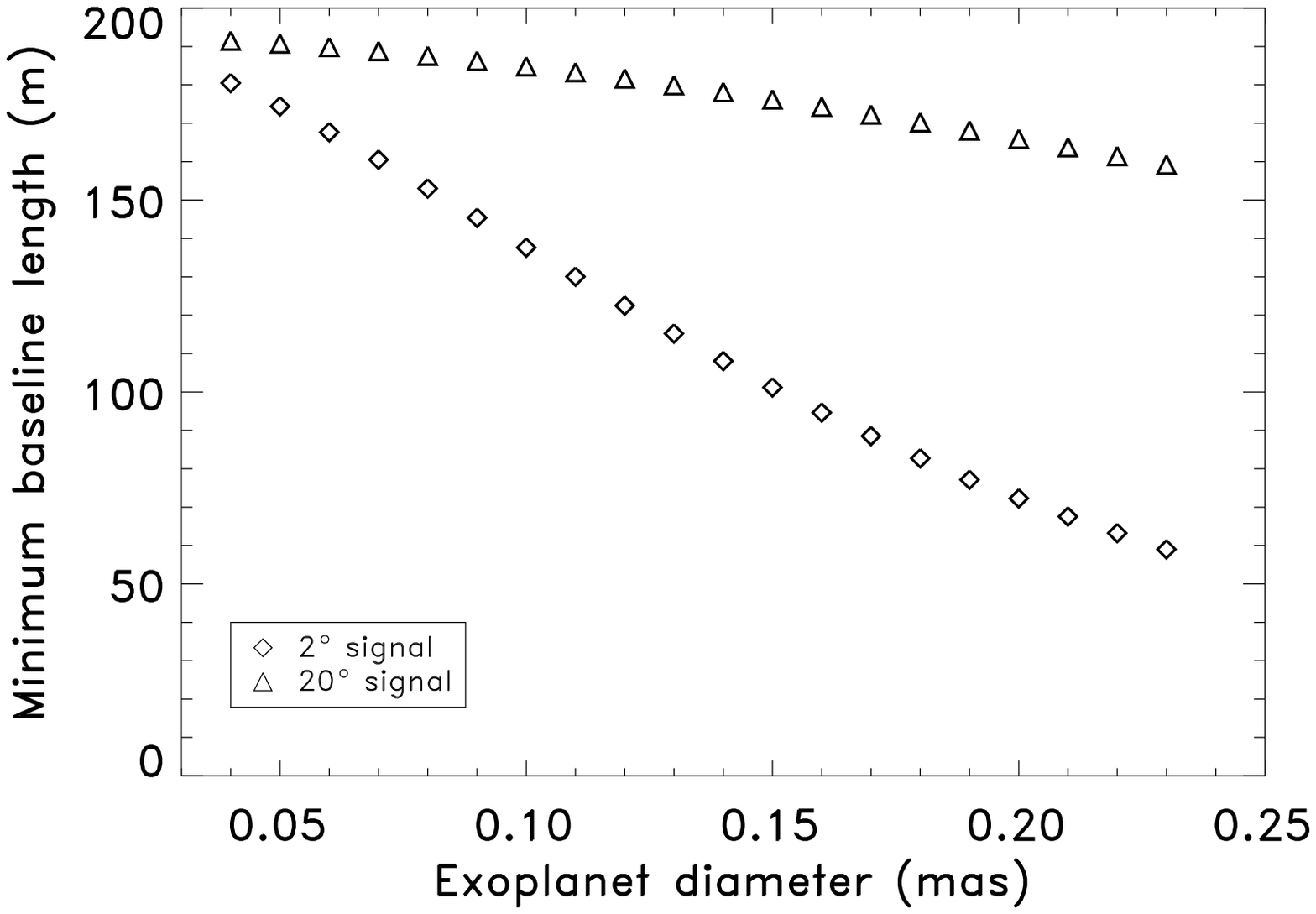} &
	 \includegraphics[scale=0.5]{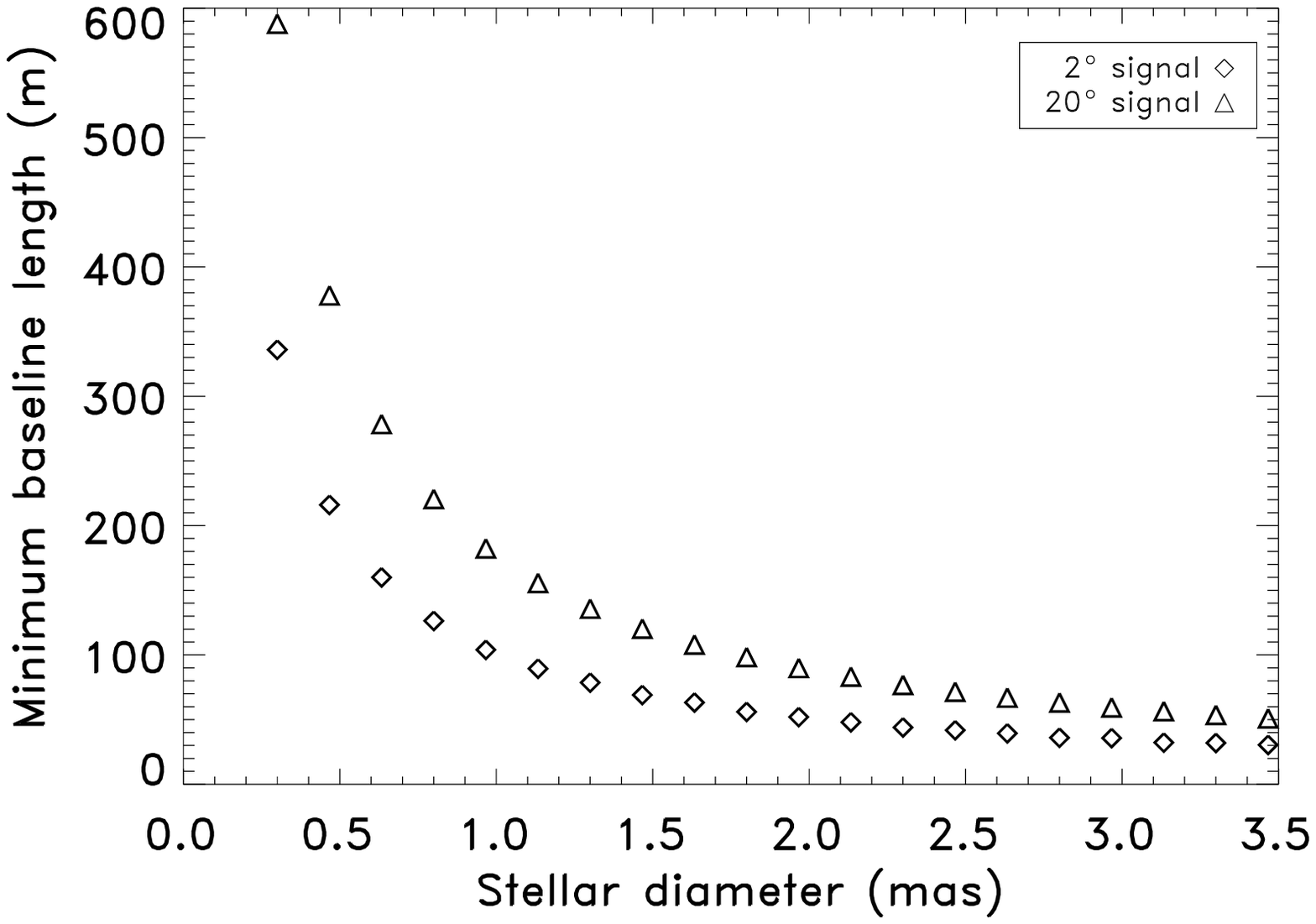} \\
	\includegraphics[scale=0.5]{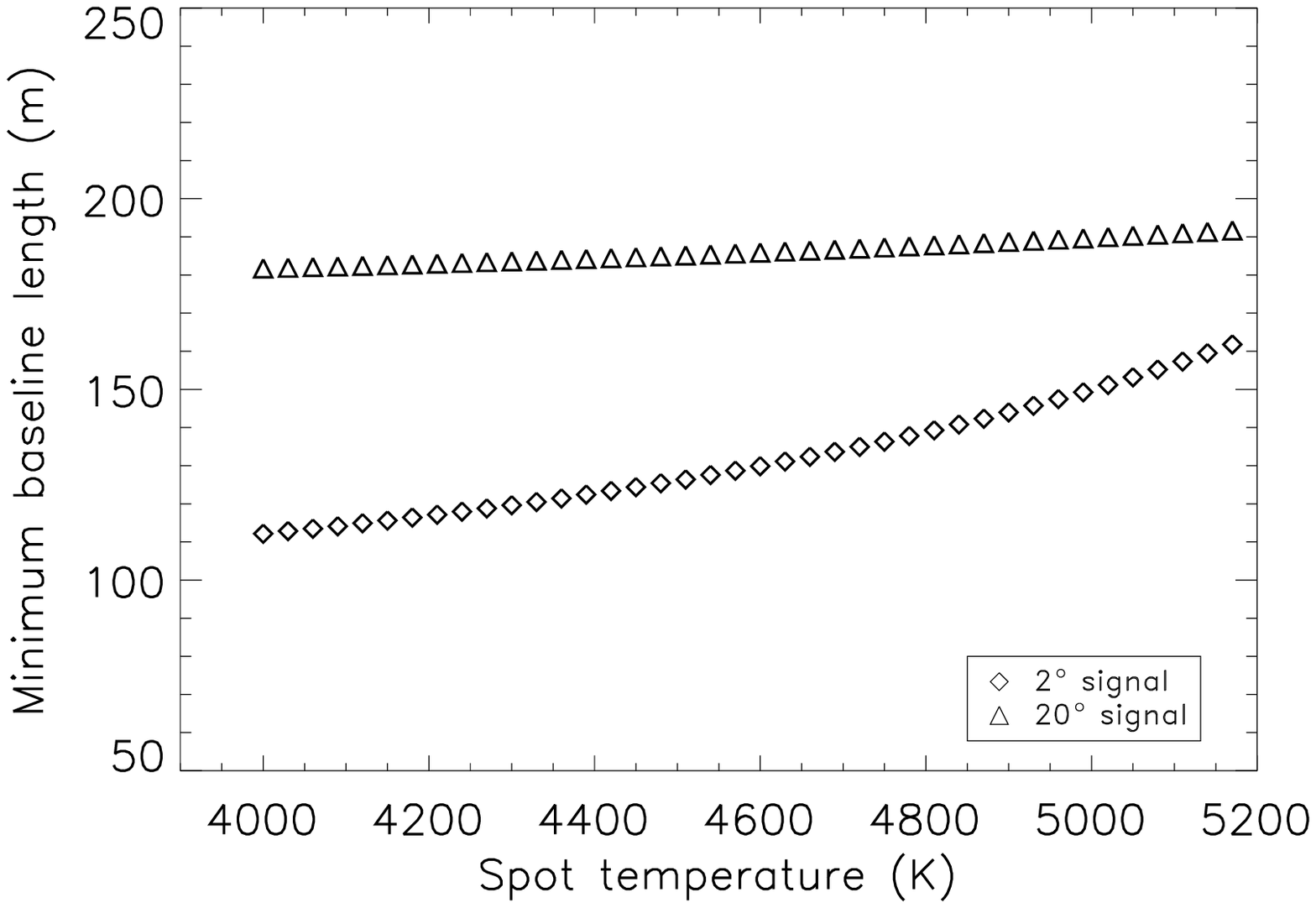} &
	\includegraphics[scale=0.5]{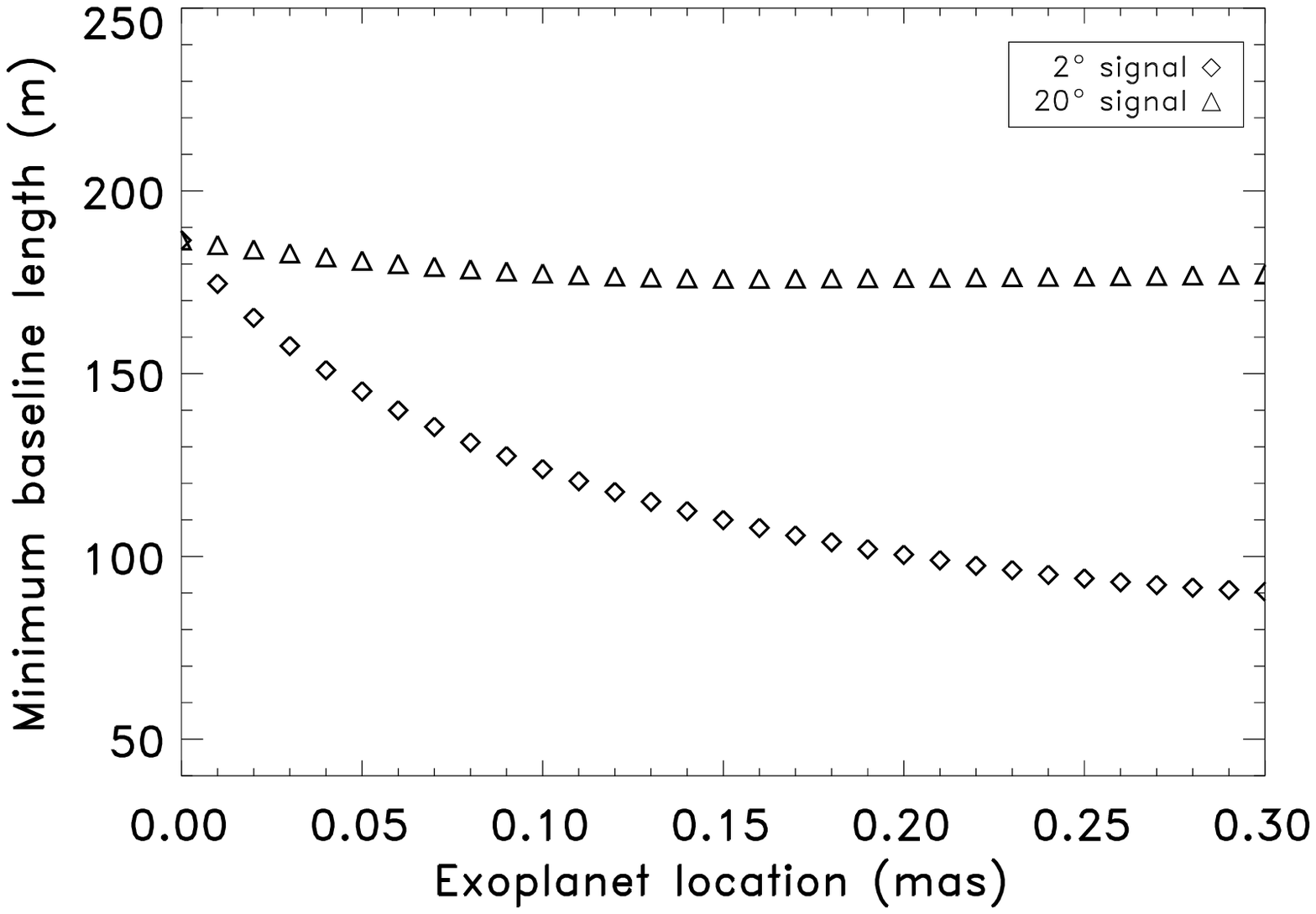}
	\end{array}$
\end{center}
\caption{Phases. Variations of the MBL according to different parameters, with the same fixed parameters as in Fig.~\ref{fig:MBLSqrdVis}. \textit{Top}: the impacting parameters are the exoplanet diameter (left) and the stellar diameter (right). \textit{Bottom}: the impacting parameters are the spot temperature (left) and the exoplanet location on the stellar disk (right), for which $\theta_{\rm p}=0.15$~mas. MBL are calculated for $20^{\circ}$ (triangles) and $2^{\circ}$ (diamonds) variations.}
\label{fig:varPhases}
\end{figure*}

\begin{figure*}
\begin{center}$
\begin{array}{cc}
\includegraphics[scale=0.5]{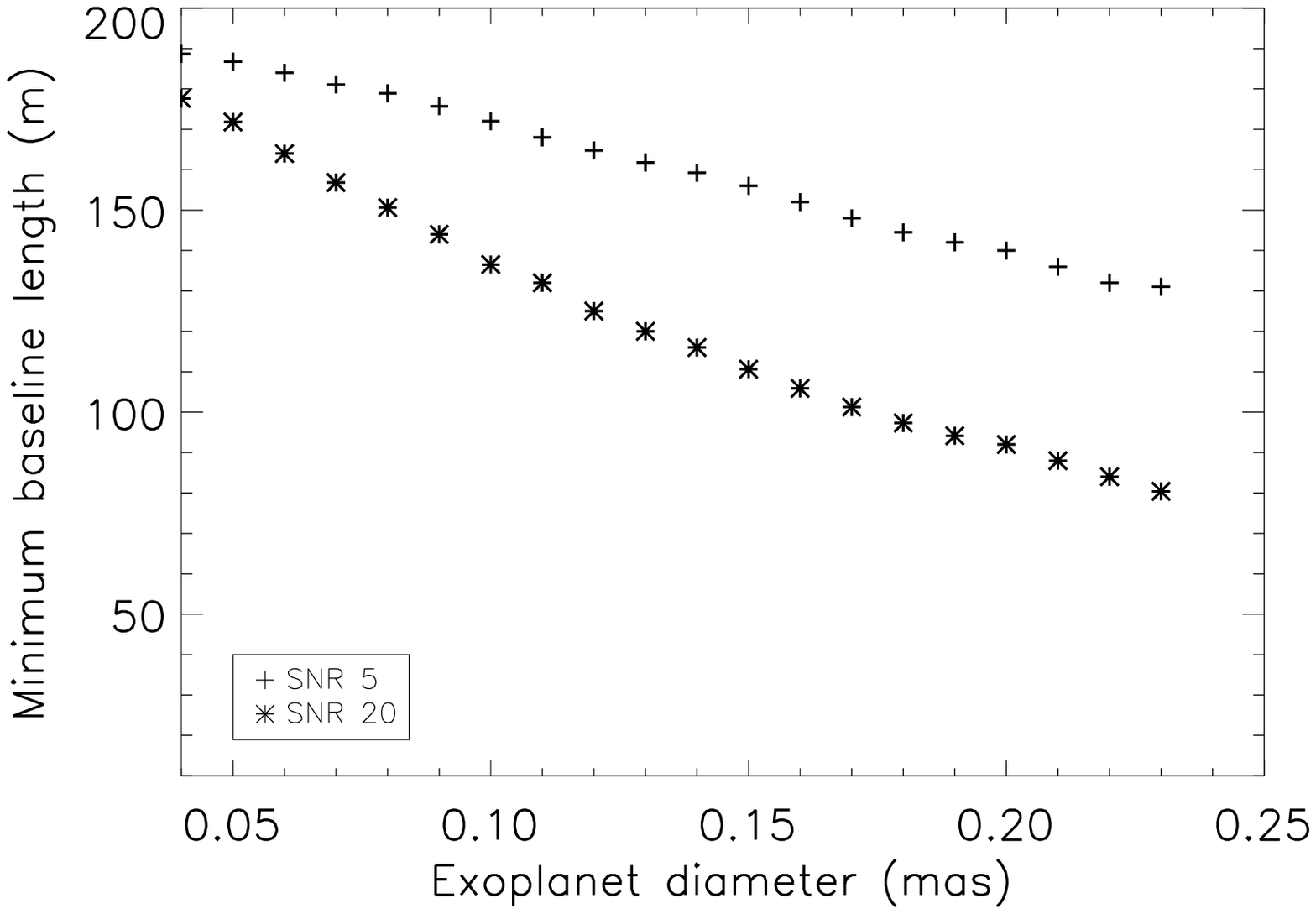} &
\includegraphics[scale=0.5]{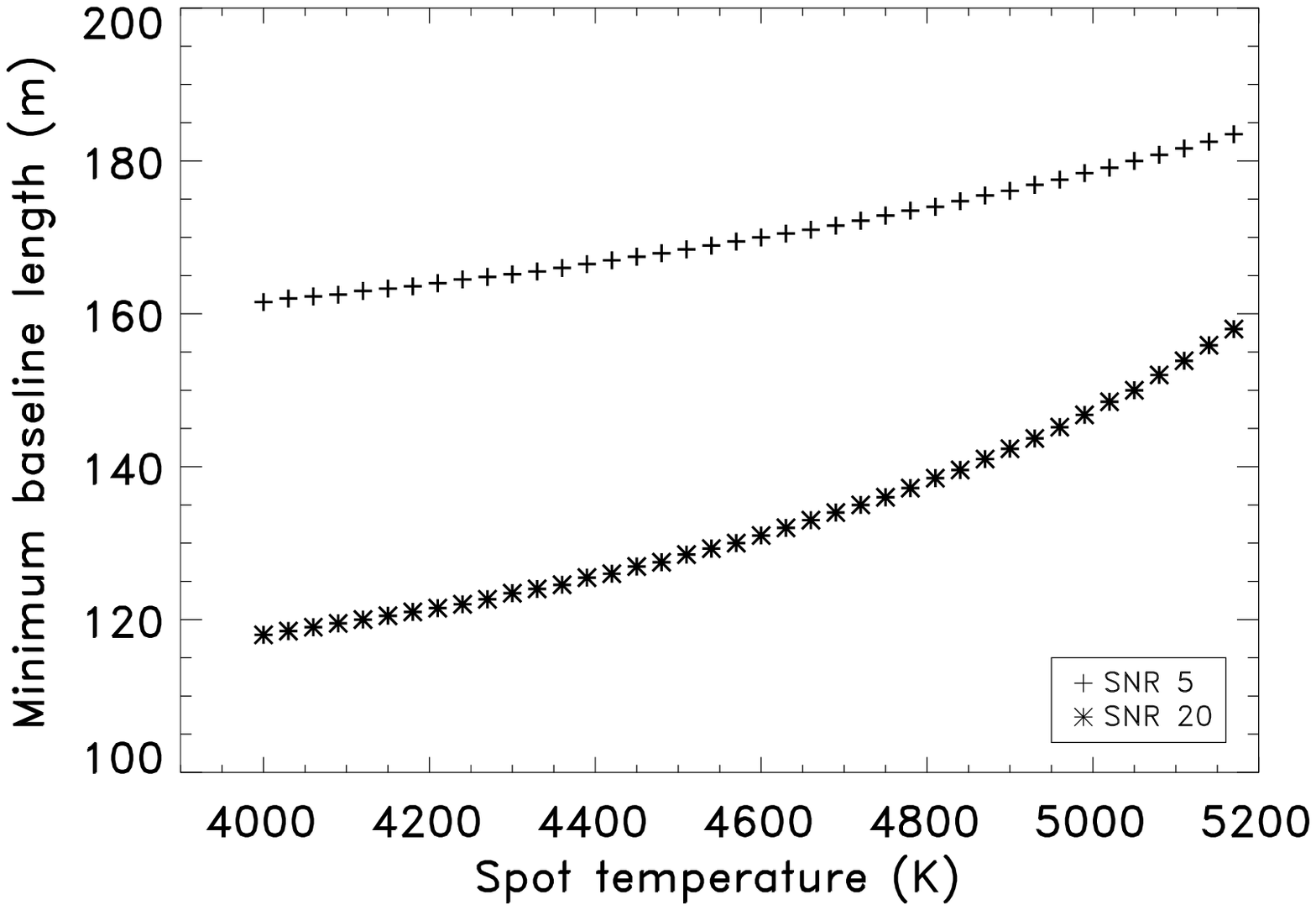}
	\end{array}$
\end{center}
\caption{Variations of the MBL considering the relative errors on squared visibilities according to the exoplanet diameter (top) and the spot temperature (bottom). We consider an S/N of $5$ (crosses) and $20$ (asterisks) and the same fixed parameters as in Fig.~\ref{fig:MBLSqrdVis}.}
\label{fig:MBLSqrdVisRel}
\end{figure*}

\subsection{Absolute or relative variation of signal}
\label{sec:Method}

We first considered the absolute variations induced by exoplanets and magnetic spots on interferometric observables, $|V^2_{\rm p+\star}-V_{\star}^2| > S$, where $V_{\star}$ is the visibility modulus of the star alone. This means that the accuracy required to detect exoplanets or spots is probably higher than the variation induced. We calculated the MBL for $S = 0.5\%, 1.0\%$, and $2.0\%$. Searching a smaller variation would be unrealistic as interferometers hardly reach this accuracy. We preliminarily calculated the MBL to identify the reachable upper limit of the signal and found no solution for better precisions than $2.0\%$. The variation of the MBL according to the different parameters is presented in Fig.~\ref{fig:MBLSqrdVis}. For the phases, we selected variations of $2^{\circ}$ and $20^{\circ}$, which are shown in Fig.~\ref{fig:varPhases}, and from which it could be possible to derive closure-phase values.

But this first approach is really conservative as it implicitly considers that the visibility measurements are not limited by photon noise, but by a fixed noise. If we consider that the signal-to-noise ratio (S/N) of a measurement is just photon-noise limited, then the uncertainty $\sigma V^2$ can be written as
\begin{equation}
 \sigma V^2 = \frac{V^2}{S/N} \ .
\end{equation}
Thus in a second step we computed the MBL for two values of the S/N: $20$ for good observing conditions and $5$ for limit conditions.
This means that we calculate the MBL as
\begin{equation}
\frac{|V^2_{\rm p} - V^2_{\star}|}{V^2_{\star}} > \frac{1}{S/N} \ .
\end{equation}
The variation of the MBL considering the different S/N are presented in Fig.~\ref{fig:MBLSqrdVisRel}.

\subsection{Results}
\label{sec:Results}

\begin{table*}[ht]
\centering
\begin{tabular}{l|  c  c  c  c  c  |c  c }
\hline \hline
					&  \multicolumn{5}{|c|}{Visibility} & \multicolumn{2}{c}{Phases} \\
					
					& $0.5 \%$  	& $1 \%$ 	 & $2 \%$ 	 & S/N $= 5$ & S/N $= 20$ 	&  $2^{\circ}$ 	& $20^{\circ}$ 	\\
\hline
$\theta_{\star}$	& $0.3$	mas		& $0.3$	mas	 &		/	  &	$0.3$	mas	& $0.3$	mas	 &	$0.3$ mas   &	$0.3$ mas	\\
					& $160$ m		& $250$ m	 &			  &	$515$ m		& $364$ m	 &	$340$ m		&	$600$ m		\\
					\hline
$\theta_{\rm p}$	& $0.09$ mas 	& $0.14$ mas & $0.20$ mas &	$0.04$ mas	& $0.04$ mas & $0.04$ mas	&	$0.04$ mas	\\
					& $110$ m		& $110$ m	 & $90$ m	  &	$190$ m		&	 $180$ m	 &	$180$ m		&	$190$ m		\\
					\hline
$T_{\rm s}$		&	$4870$~K		&	$4030$~K &		/	  &	$5170$~K		&	$5170$~K	 &	$5170$~K	&	$5170$~K		\\
					&	$100$ m		&	$108$ m	 &			  &	$175$ m		&	 $158$ m	 &	$184$ m		&	$192$ m		\\
					\hline
$\alpha_{\rm p}$	& 		C	&		C	 &		C	  &		C		&	C	 &	$0.3$ mas	&	$0.3$ mas	\\
						&	$46$m	&	$73$ m	 &	$250$ m	  &		$155$ m	&	 $110$ m	 &	$90$ m		&	$180$ m		\\
\hline
\end{tabular}
\caption{Summary of the results obtained for the MBL tests. For each tested parameter we list the lowest or highest value of our test for which a MBL is found and the value of the corresponding MBL (see text). C stands for constant. }
\label{tab:Summary}
\end{table*}

Table~\ref{tab:Summary} is a summary of the main results. It describes the most interesting values of the MBL obtained for the different parameters. Thus, it gives the MBL obtained for the smallest tested exoplanet diameter, stellar diameter, highest tested spot temperature, and the MBL corresponding to the exoplanet position. For this last parameter, an average of the MBL is given since it is almost constant.

The analytic expression of the complex visibility is a function of $z = \pi \theta_\star B/\lambda$ only, so we expect the MBL to be proportional to $\theta_\star^{-1}$ (and to $\lambda$). This is indeed the case, as shown in the upper panel of Fig.\ref{fig:MBLSqrdVis}, where the data points are given by \texttt{COMETS} and the lines are $76/ \theta_\star$ (solid line) and $47/ \theta_\star$ (dashed line). The same behavior is observed in the top right panel of Fig.~\ref{fig:varPhases}, where we see that baselines longer than $350$~m are necessary to detect the planet in phases. The MBL are slightly higher for a spot than for an exoplanet (for example, for a $4300$~K spot, we find that MBL of $401$~m and $604$~m are necessary to obtain signals of $2^{\circ}$ and $20^{\circ}$ for $\theta_\star = 0.3$~mas).

Spots of temperature $T=5170$~K (the highest tested value) can also be detected in phases and considering the two chosen S/Ns. However, there is no detectable absolute signal of $2 \%$ due to the spot, and there are solutions only for $T_{\rm s}<4030$~K, which defines the upper limit to obtain an absolute signal of $1 \%$ in this study.

The exoplanet diameter has a strong impact on the MBL. Even under poor conditions (S/N = 5), a small exoplanet ($\theta_{\rm p} = 0.04$ mas) can be detected with hectometric baselines. However, there is no detectable absolute variation of the squared visibility for small exoplanets ($\theta_{\rm p} < 0.09$~mas for a signal of $0.5 \%$).

According to this model, baselines that already exist should be long enough to detect exoplanetary or spot signals in visibilities if the exoplanet is large enough or the spot dark enough (with a low temperature). To detect small exoplanets or hotter spots, phase measurements are more appropriate as their signals reach low but still detectable values (like $2^{\circ}$). For the same diameter, the MBL needed to detect a spot is larger than for an exoplanet because of its contrast with the star, which is stronger for the exoplanet. An exoplanet produces a higher signal on interferometric observables that is thus easier to detect.  For large enough stars ($\theta_{\star} > 0.3$ mas), CHARA baselines are long enough even now to detect a certain variation of the interferometric observables caused by exoplanets or spots. The problem lies in instrument accuracy (see next section).

Finally, from these results, one can easily derive a general empirical formula to compute the MBL. The difference between the reference and perturbed squared visibilities reads:
\begin{equation*}
\left| \left( \frac{\tilde{I}_{\star} + \tilde{I}_{\rm p}} {\tilde{I}_{\star}(0) + \tilde{I}_{\rm p}(0)} \right)^2 -  \left( \frac{\tilde{I}_{\star}} {\tilde{I}_{\star}(0)} \right)^2 \right|\ .
\end{equation*}
The perturbing body can be detected if, and only if, this difference is larger than $S$. Expanding to first order, this is equivalent to
\begin{equation}
\frac{\theta_{\rm p}}{\theta_{\star}} < \sqrt{\frac{S}{2}} \ .
\end{equation}
This means that the exoplanet angular diameter has to be larger than $\theta_{\rm min} = \sqrt{\frac{S}{2}} \times \theta_{\star}$ to be detected. Therefore, we can look for a general formula of the form $MBL = \Gamma(\theta_\star, S, \lambda) \left( \frac{\theta_{\rm p}} {\theta_{\star}} - \sqrt{\frac{S}{2}} \right)^{\Delta(S)}$. As we have seen in Sec.~\ref{sec:Results}, $\Gamma$ must be proportional to $\lambda/\theta_\star$. We find the dependence of $\Gamma$ and $\Delta$ on $S$ by applying a least-squares method to find that the best fit is
 \begin{equation}
 \begin{aligned}
 	MBL &= (17\times S^{0.1}) \times \left( \frac{\theta_{\rm p}} {\theta_{\star}} - \sqrt{\frac{S}{2}} \right)^{-1.7\times S^{0.2}}  \\
 	&  \times \left(\frac{\theta_{\star}}{1 \rm mas}\right)^{-1} \rm \times \frac{\lambda}{720.10^{-9}}\  \rm m \ .
 \end{aligned}
 \label{equ:EquationGenerale}
\end{equation}
This law is plotted in Fig.~\ref{fig:MBLSqrdVis} (solid, dashed, and dotted lines, middle panel). It gives a good fit of the MBL.

As an example, we have taken the host star HD$189733$, whose angular diameter is $\theta_{\star} = 0.38$ \citep{Baines2008} and whose transiting exoplanet angular diameter is $\theta_{\rm p} = 0.056$~mas, derived from the combination of its linear radius \citep{Torres2008} and the stellar distance \citep{vanLeeuwen2007}. At $\lambda = 720$~nm, we find with the formula $MBL = 218$~m for a signal of $1 \%$, and $MBL = 454$~m for a signal of $2 \%$. As expected, since the angular resolution is $\propto B/\lambda$, with $B$ the baseline, these values are much lower than in the infrared domain. For the same star, signals of $1 \%$ and $2 \%$ in the K band ($2.13 \rm \mu$m) would be reached with $MBL = 645$~m and $MBL = 1344$~m.
Since we only considered transiting exoplanets, the difference of contrast (visible versus infrared) between an exoplanet and its host star does not change the results.

\subsection{Discussion}
\label{sec:Detectability}

\begin{figure*}
\begin{center}$
\begin{array}{cc}
	\includegraphics[scale=0.50]{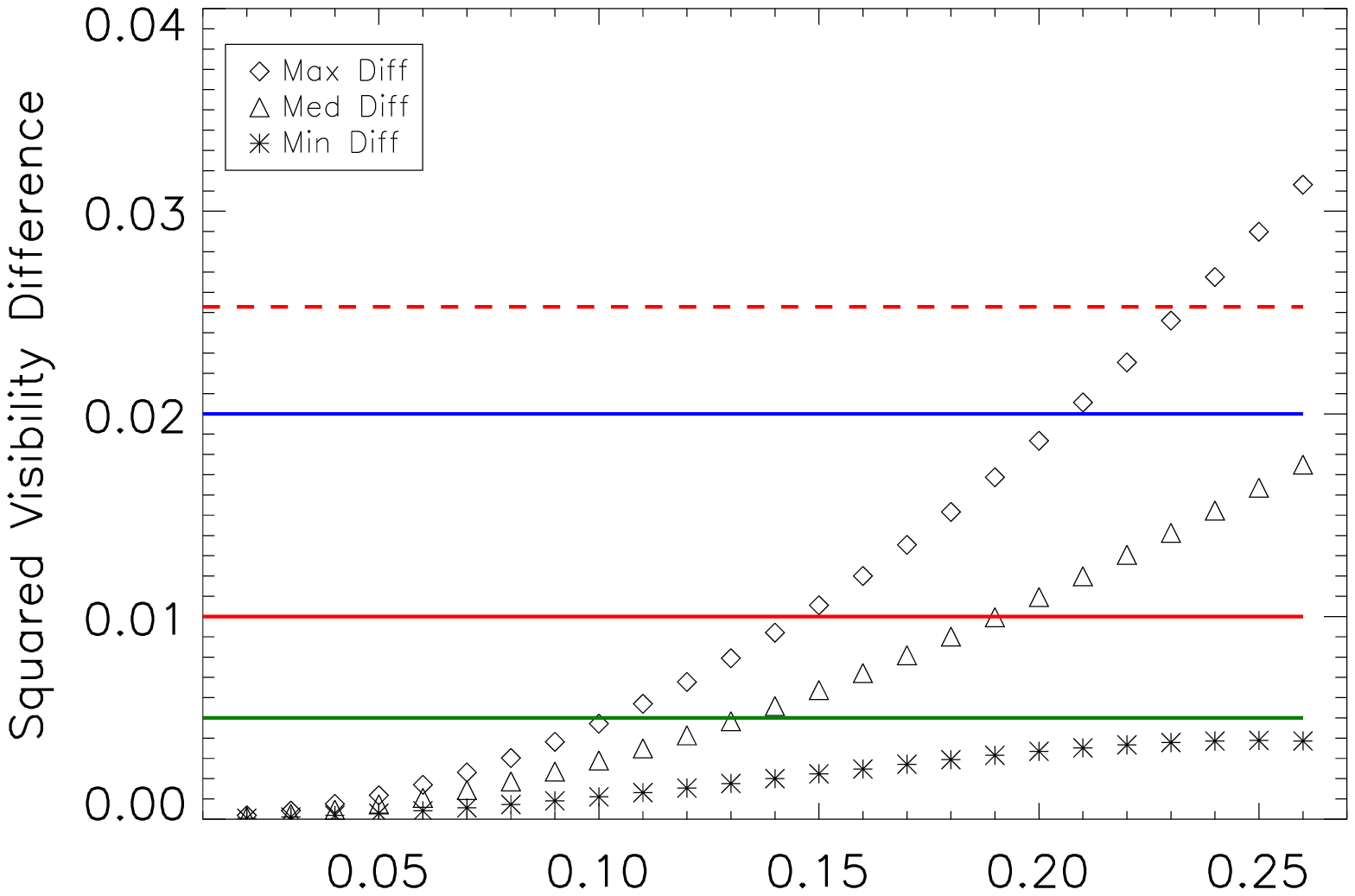} &
	 \includegraphics[scale=0.50]{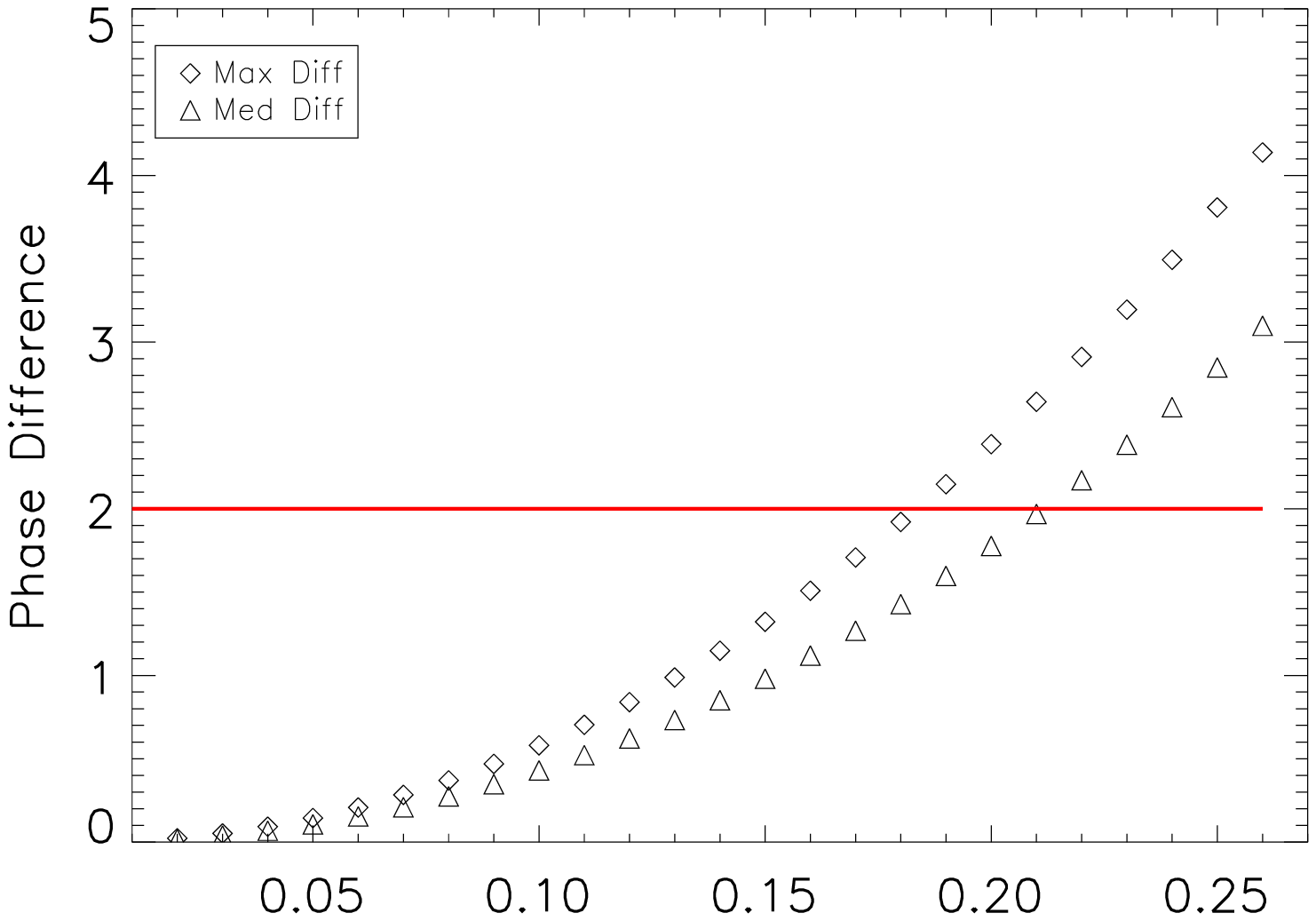} \\
	\includegraphics[scale=0.50]{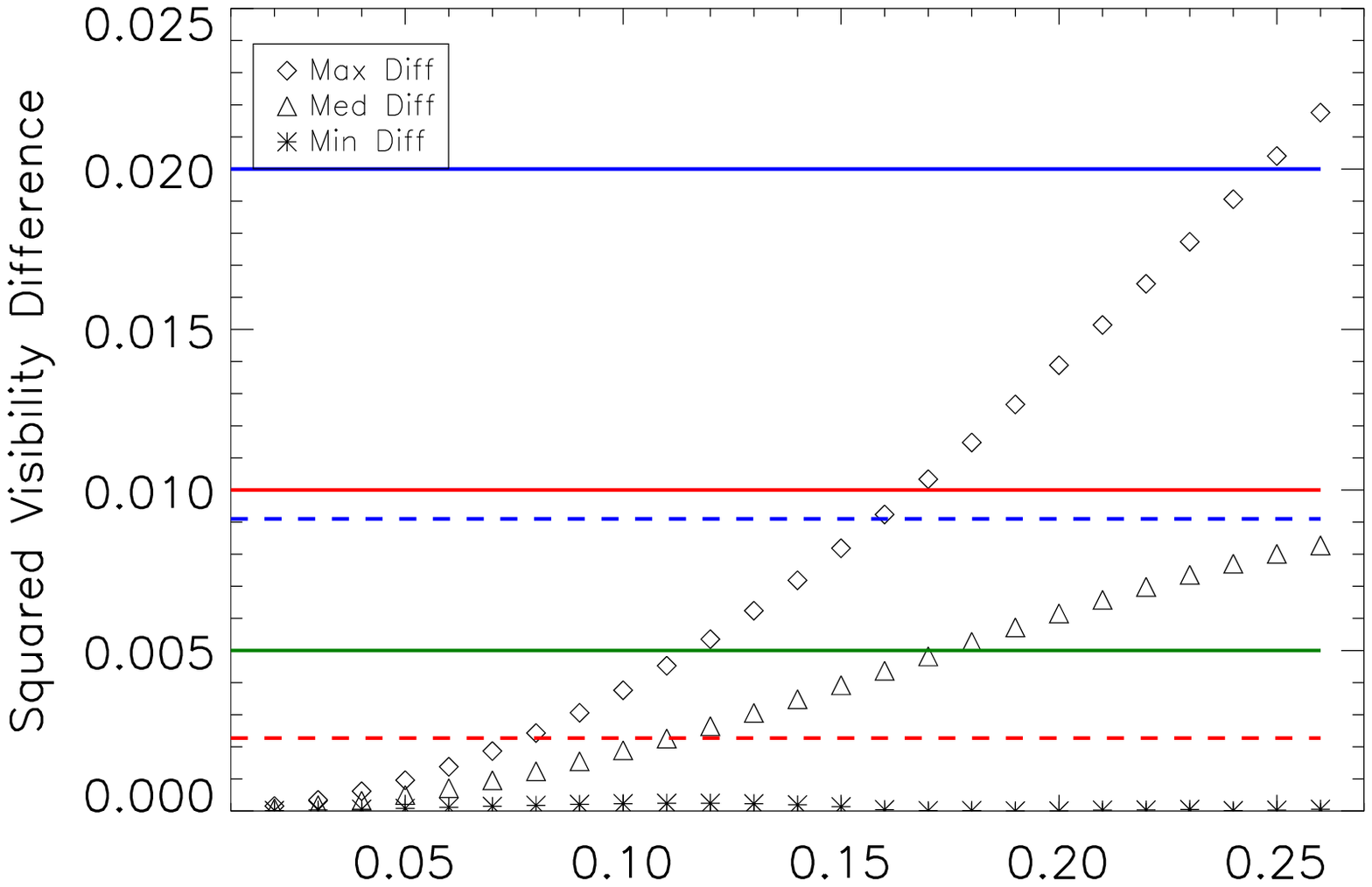} &
	 \includegraphics[scale=0.50]{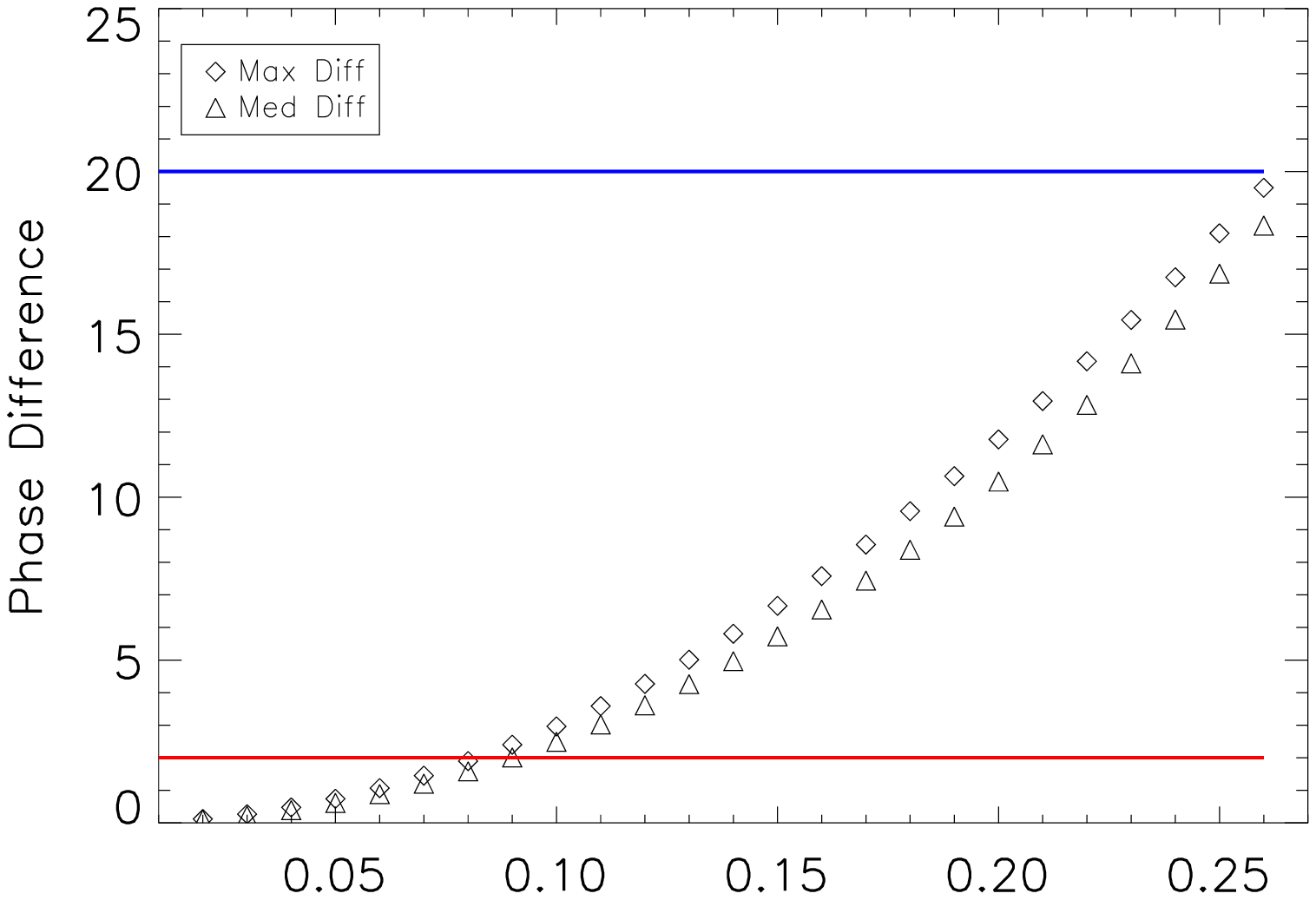} \\
	\includegraphics[scale=0.50]{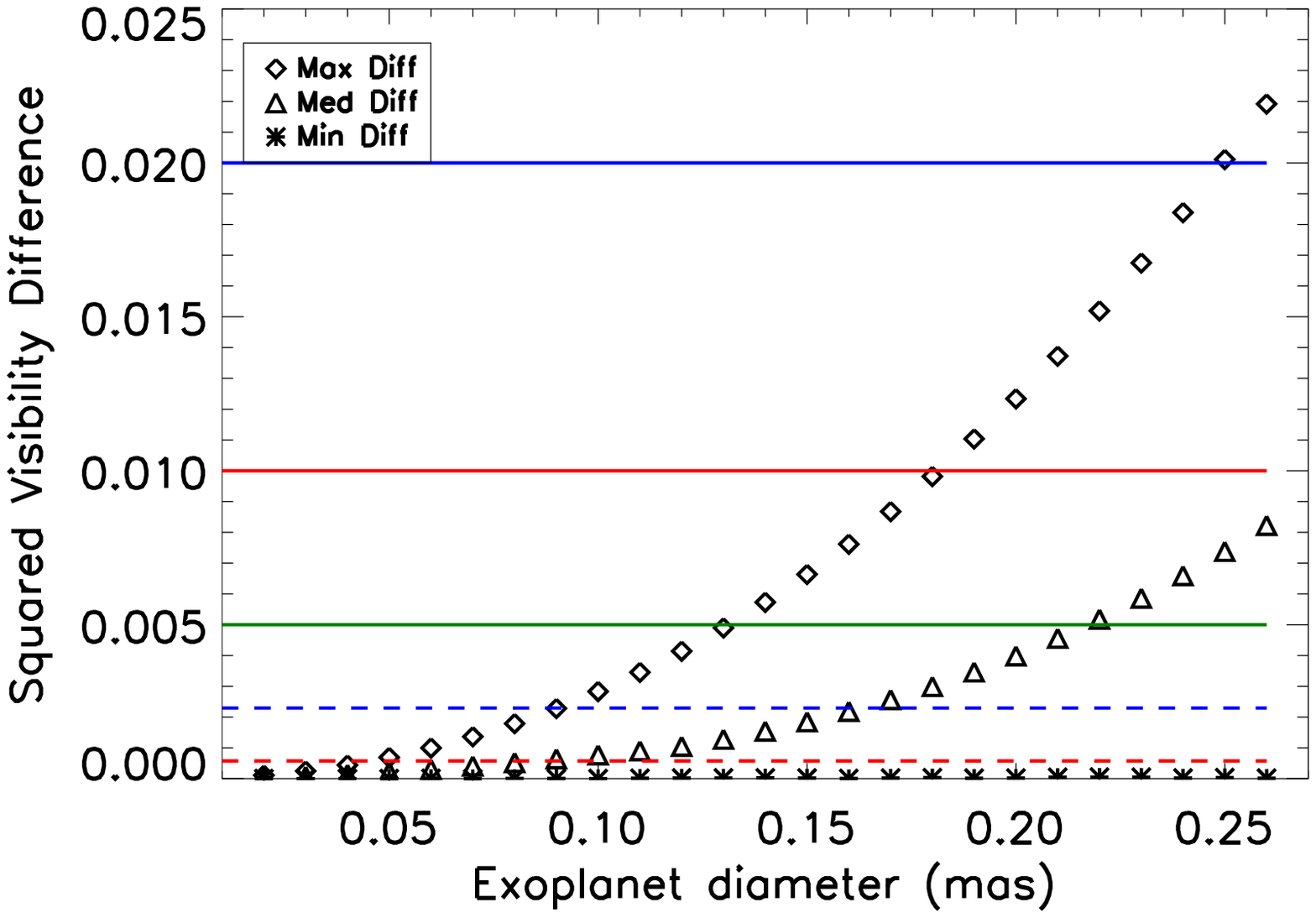} &
	 \includegraphics[scale=0.50]{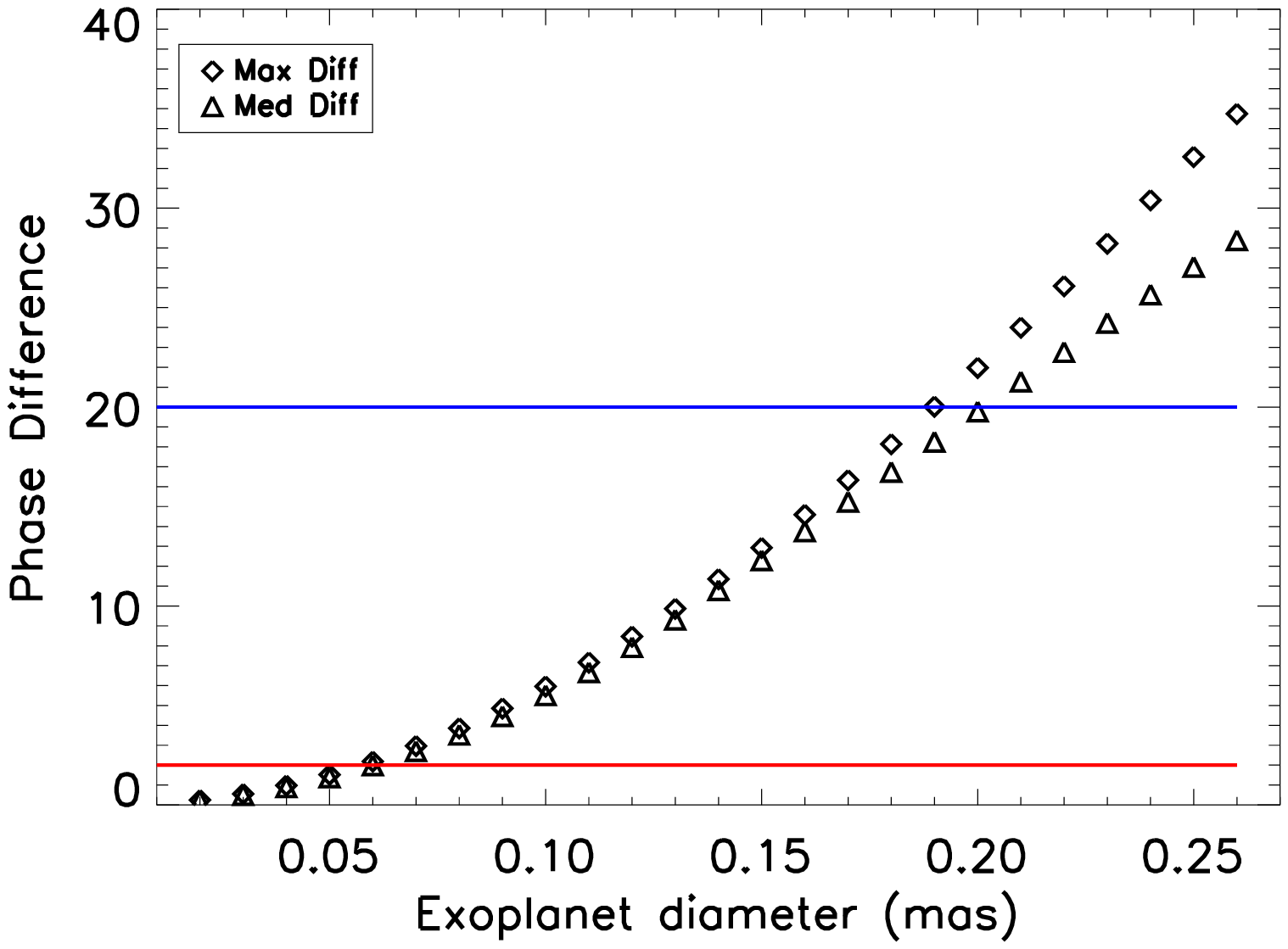} \\	
\end{array}$
\end{center}
\caption{Squared visibility and phase differences between a 1~mas star with a transiting exoplanet at ($0.2, 0.0$)~mas and a single star at different spatial frequencies ($1.1\times10^{8}$ ($1$/rad), top; $2.1\times10^{8}$ ($1$/rad), middle; $3.4\times10^{8}$ ($1$/rad), bottom). Squared visibilities: \textit{Solid lines}: Absolute errors ($1\%$, red; $2\%$, blue; $0.5\%$, green). \textit{Dashed lines}: Errors corresponding to different S/N: $5$ (blue) and $20$ (red). Phases (in degree): $2^{\circ}$ (red) and $20^{\circ}$ (blue).}
\label{fig:Seuils}
\end{figure*}

Knowing at which spatial frequency (and thus in which visibility lobe) an exoplanet or spot signal is detectable will bring information on the sensitivity limit that interferometers have to provide to make such measurements, including phase and closure-phase measurements. For three different spatial frequencies corresponding to the middle of the first lobe of visibility ($1.1\times10^{8}$ $1$/rad), close to the first null ($2.1\times10^{8}$ $1$/rad), and in the middle of the second lobe ($3.4\times10^{8}$ $1$/rad), the largest, smallest and median differences between the squared visibility of a transited star and a single star are computed, along with the largest and median differences in phase (the smallest difference is zero and is not represented). They are tested according to the exoplanet diameter because this parameter induces strong variations of the MBL, as shown in Sect.~\ref{sec:GeneralStudy}. They include the complete ($u, v$) plane. Figure~\ref{fig:Seuils} shows for each of these three cases the level corresponding to $2 \%$, $1 \%$, and $0.5 \%$ absolute differences (solid lines) and relative variations for an S/N of $20$ and $5$ (dashed lines). It also provides the levels $2^{\circ}$ (red) and $20^{\circ}$ (blue) for phases.

In the first visibility lobe, the highest difference between two visibilities ($2 \%$) occurs for $\theta_{\rm p} > 0.20$~mas, which does not correspond to an exoplanet diameter but rather a brown dwarf. Even a good S/N cannot provide a sufficient signal to detect an exoplanet (solution of $\theta_{\rm p} > 0.22$~mas for the red dashed line). However, there is a highest absolute signal of $0.5 \%$ and $1 \%$ on squared visibilities for exoplanet diameters such as $\theta_{\rm p} > 0.10$~mas and $\theta_{\rm p} > 0.15$~mas which correspond to real cases. The signal on the phase measurement only reaches $2^{\circ}$ for $\theta_{\rm p} > 0.17$~mas, which is almost the highest limit for a planet representation.

The visibilities show larger differences close to the zero of visibility and in the second lobe when considering relative errors. For an S/N=20, very small exoplanets ($\theta_{\rm p} > 0.08$~mas for the smallest) generate a variation of the visibility that can be detected close to the zero of visibility, and exoplanets as small as $\theta_{\rm p} = 0.04$~mas generate variations that can be measured in the second lobe. In poor conditions (S/N=5), small exoplanets can still be detected. The smallest ones are such that $\theta_{\rm p} > 0.16$~mas and $\theta_{\rm p} > 0.10$~mas for the first zero of visibility and in the second lobe. This shows that interferometry is a real opportunity for detecting exoplanets.

Absolute variations of the visibility exist close to the zero and in the second lobe, but are found for larger planets compared to relative errors. To obtain a $1 \%$ signal, the exoplanet diameter has to be $\theta_{\rm p} > 0.17$~mas. However, a highest signal of $0.5 \%$ is detectable for only $\theta_{\rm p} > 0.13$.

Thus, if focusing on relative differences, which take into account the S/N and thus reproduce a real observing condition, signals of small exoplanets can be identified near the first zero of visibility and in the second lobe. For each frequency, they never allow detecting very small exoplanets ($\theta_{\rm p} < 0.05$~mas) for these considered signals. The instrument sensitivity is thus a fundamental parameter to be improved.

The S/N is lower when measuring low visibilities, for instance close to the first zero of visibility and in the following lobes. This is where the higher order measurements (phase or closure phase) show evidence of the exoplanet or spot, however, that is a highest signal. In our study, the phase signal reaches $2^{\circ}$ for a star with an exoplanet of diameter $\theta_{\rm p} > 0.08$~mas and $\theta_{\rm p} > 0.06$~mas close to the zero and in the second lobe. This is also where planetary companions are able to cause a phase variation of $20^{\circ}$, but this only concerns large planets ($\theta_{\rm p} > 0.25$ and $\theta_{\rm p} > 0.17$~mas).
Thus, taking the S/N into account tells us whether the signal of the exoplanet or star can also be measured by these higher order observables.

\begin{table}
\begin{tabular}{llll}
\hline \hline
Instrument 	 &  $V^2$ accuracy 	& CP accuracy	& Ref. \\
\hline
VEGA/CHARA	 &  $1-2 \%$		& 	-			& 1 \\
FLUOR/CHARA	 &  $0.3 \%$		& 	-			& 2 \\
JouFLU/CHARA &	$0.1 \%{^\ast}$	&	-			& 3 \\
			 &  $1 \%$ 			&	-			& 4 \\
VISION/NPOI	 &$5-20 \%^{\ast\ast}$& $1-10^{\circ}$& 5 \\
CLIMB/CHARA	 &	$5 \%$ 			& $0.1^{\circ}$	& 6 \\
CLASSIC/CHARA&  $5 \%$			&	-			& 7 \\
PAVO/CHARA	 &	$\sim 5\%$		& $5^{\circ}$	& 8 \\
MIRC/CHARA	 & 	$\sim 2 \%$		& $< 1^{\circ}$	& 9 \\
			 &		-			&  $0.1-0.2$	& 10, 11, 12 \\
AMBER/VLTI	 &	-				& $0.20-0.37$	& 13 \\
PIONIER/VLTI &	-				& $0.25-3^{\circ}$	& 14 \\
			 & $3-15 \%$		& $0.5^{\circ}$ & 15 \\
GRAVITY/VLTI &		-			& $1$				& 16 \\
MATISSE/VLTI &	$1.6-2.3\%$		& $<1.16$			& 17 \\
\hline
\end{tabular}
\caption{Summary of the accuracy reached on squared visibilities $V^2$ and closure phases ($CP$) by different instruments.\\
$^\ast$Expected in visibility modulus $^{\ast \ast}$In visibility amplitude.\\
Ref.: $^1$\cite{Mourard2009} $^2$\cite{Coude2003} $^3$\cite{Scott2013} $^4$\cite{Lhome2012} $^5$\cite{Garcia2014}, evolving to accuracies of $<10\%$ and $<5^{\circ}$ in $V^2$ and CP (E. Garcia, private communication) $^6$\cite{tenBrummelaar2012} $^7$\cite{McAlister2012} $^8$\cite{Maestro2012} $^9$\cite{Che2012} $^{10}$\cite{Zhao2011} $^{11}$\cite{Zhao2010} $^{12}$\cite{Zhao2008} $^{13}$\cite{Absil2010} $^{14}$\cite{Absil2011} $^{15}$\cite{LeBouquin2011} $^{16}$\cite[][Final Design Review 2011, private communication]{Chiavassa2014} $^{17}$\cite{Lopez2012}.}
\label{tab:PrecisionsInstrumentales}
\end{table}

Table~\ref{tab:PrecisionsInstrumentales} displays the accuracy reached by several instruments on CHARA, NPOI, and the VLTI, which can be found in the literature or were extracted from \cite{Chiavassa2014}. It is difficult to link instrumental precisions, which for most of them depend on particular cases, with a general conclusion on the feasability of exoplanet characterization. However, comparing Table~\ref{tab:PrecisionsInstrumentales} with the information provided by Fig.~\ref{fig:Seuils}, we note that
\begin{itemize}
\item for most of instruments, the accuracy on the squared visibility is not sufficient to detect small exoplanets ($\theta_{\rm p}<0.10$~mas under good observing conditions). JouFLU could be a good candidate, but the expected $0.1\%$ accuracy is not enough in its wavelength range (K' band). Instrument accuracy needs to improve by a factor 10 to be able to characterize exoplanets.
\item the accuracy on CP is generally reached with measurements on the first lobe of visibility and cannot be measured beyond (because of the wavelength or instrument sensitivity). If this accuracy is reached for measurements beyond the first null, then an exoplanet characterization could be achieved.
\end{itemize}

\section{Detecting a transiting planet in presence of stellar activity noise}
\label{sec:Detectionwithactivity}

The impact of an exoplanet and a spot on interferometric observables has been discussed separately, and the impacting parameters are now known. We now wish to explore the effect of a spot on the characterization of a transiting exoplanet.  

\begin{figure*}
\begin{center}$
\begin{array}{ccc}
\includegraphics[scale=0.40]{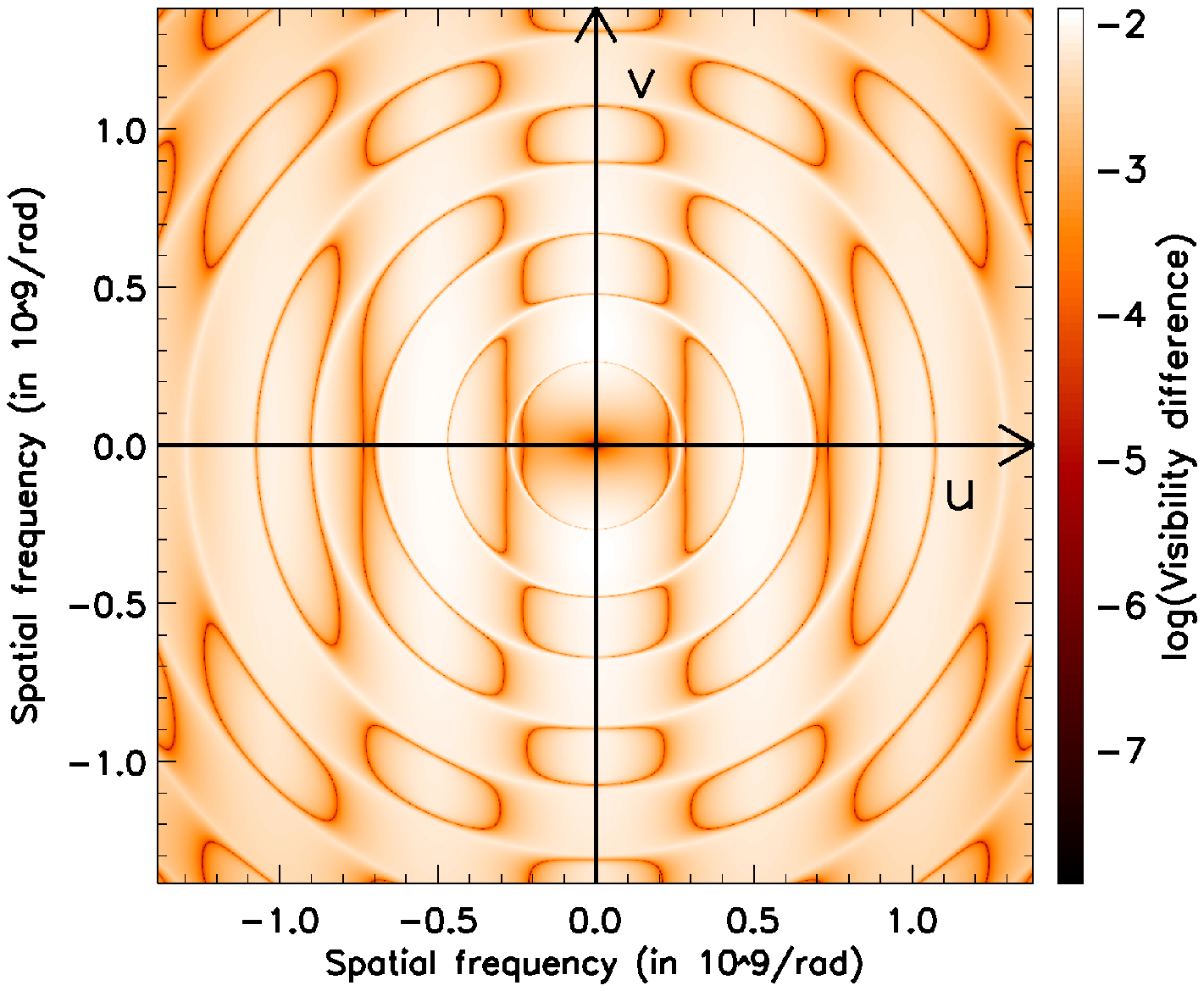} &
\includegraphics[scale=0.40]{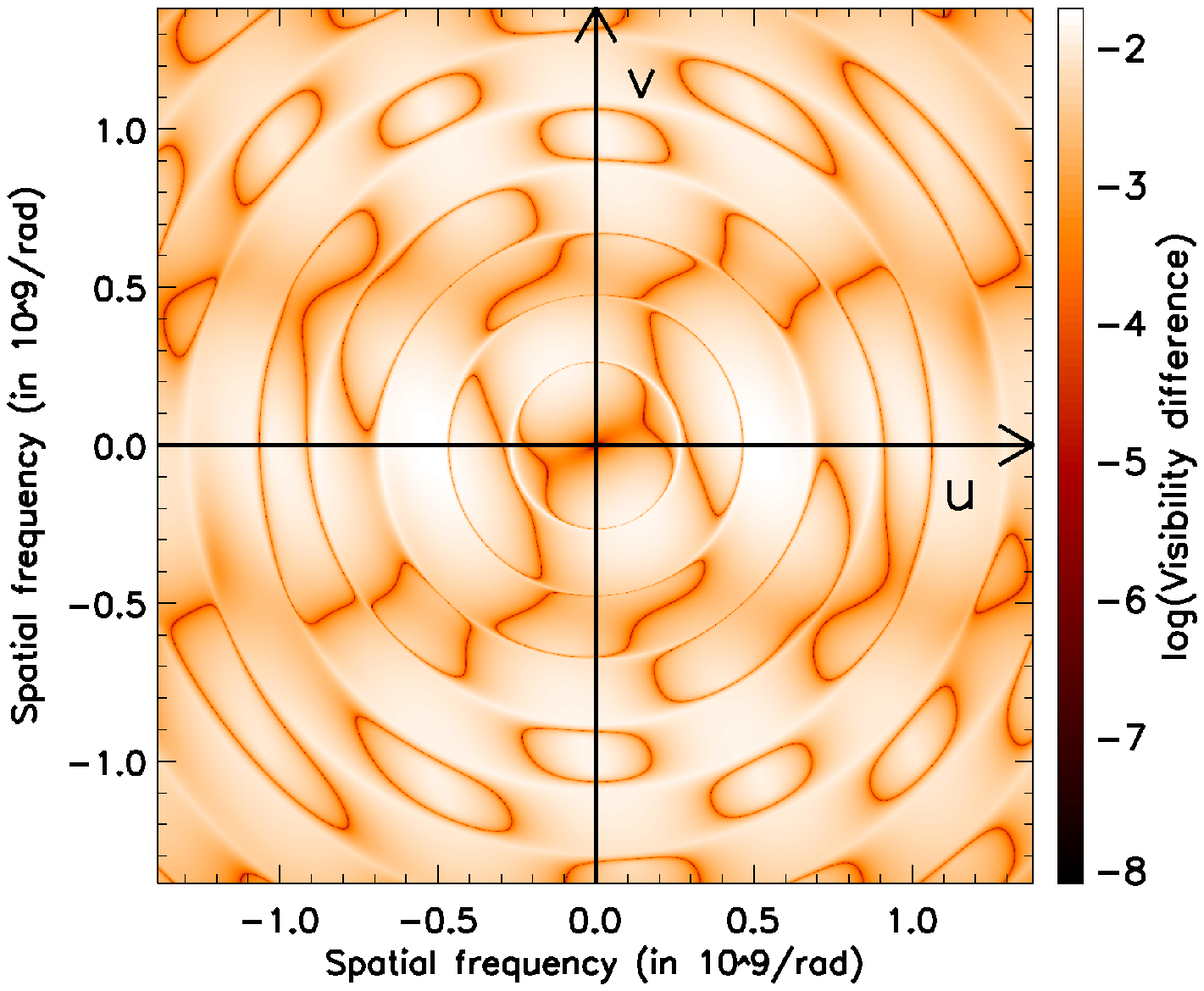} & 
\includegraphics[scale=0.40]{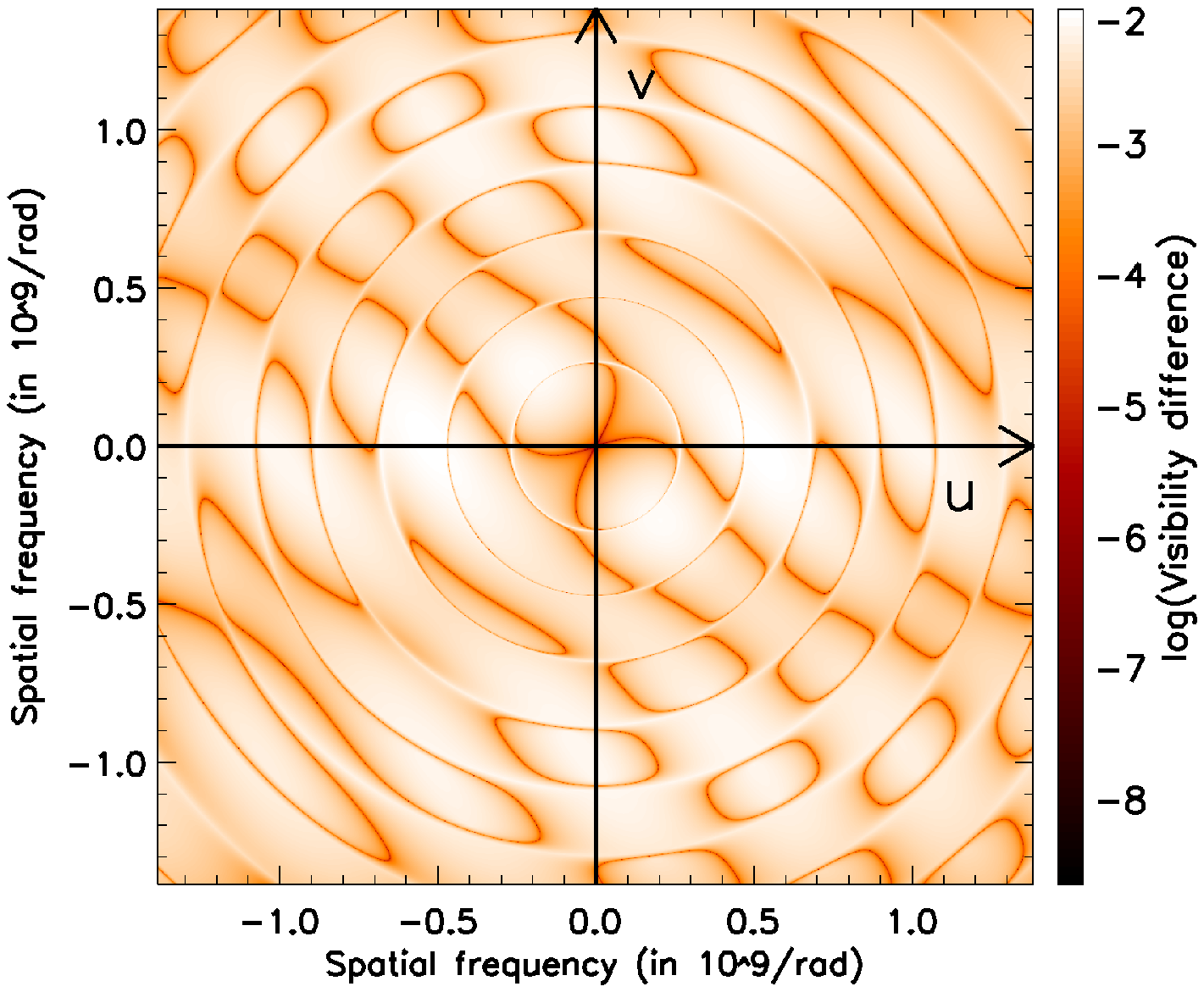} \\
\includegraphics[scale=0.40]{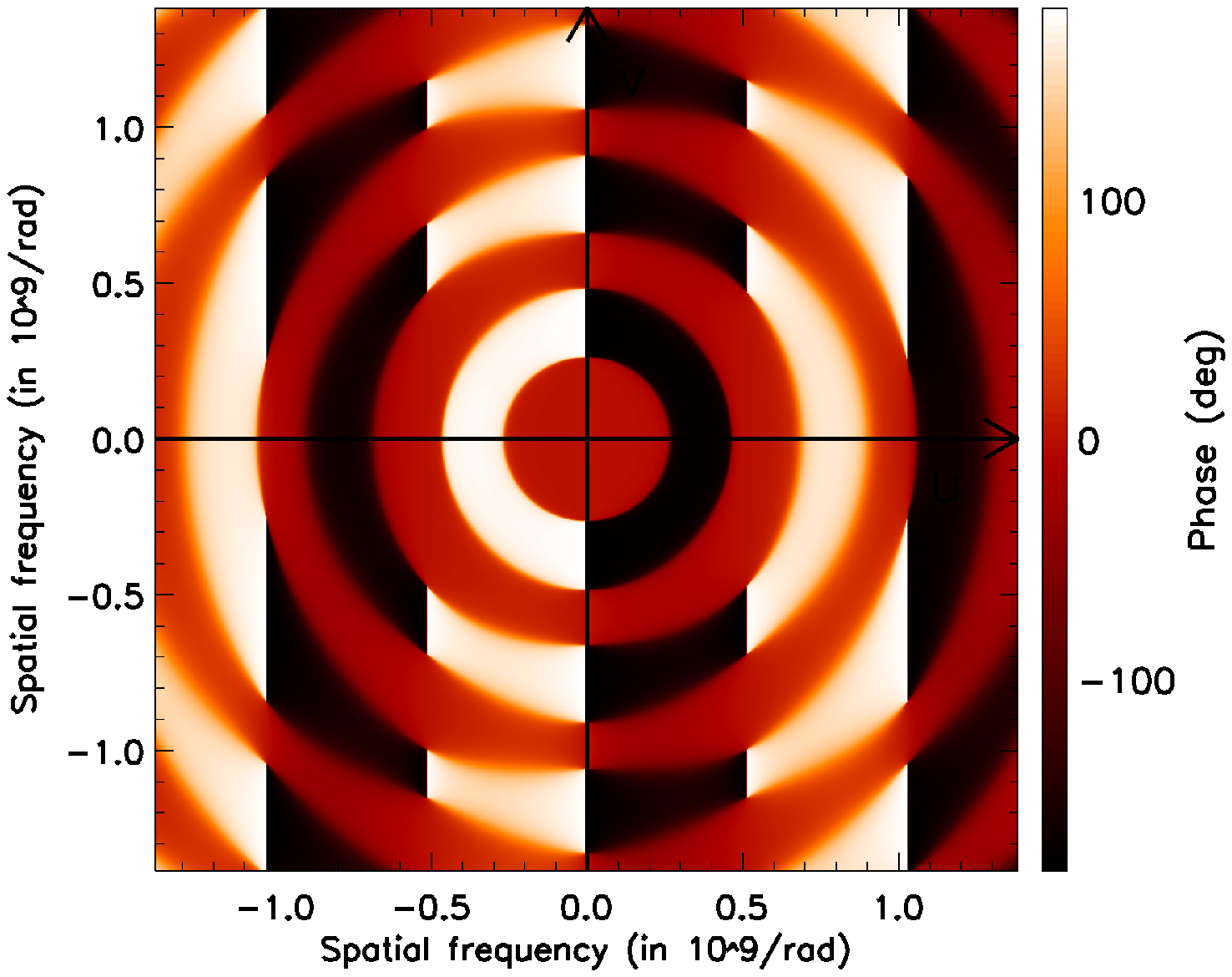} &
\includegraphics[scale=0.40]{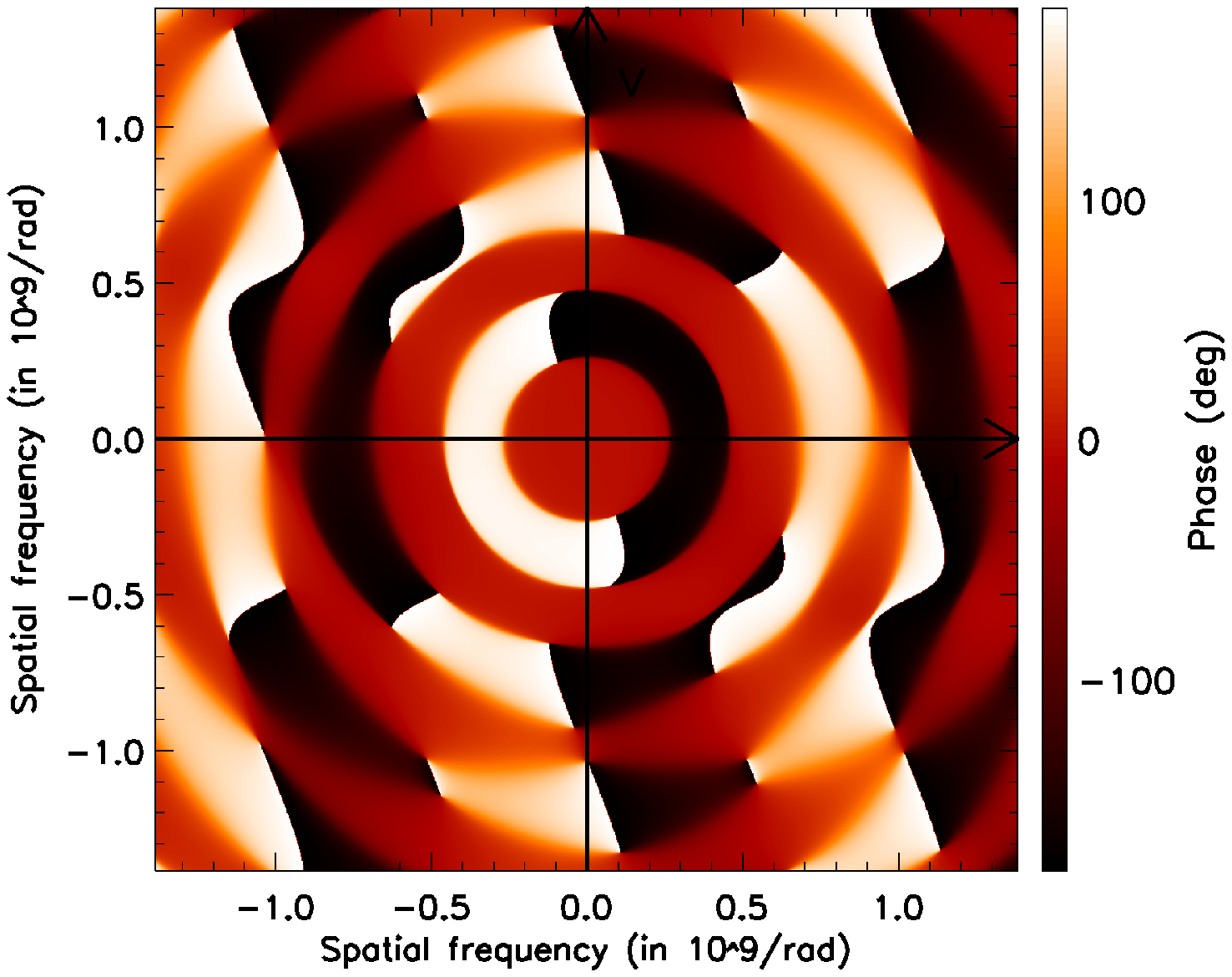} & 
\includegraphics[scale=0.40]{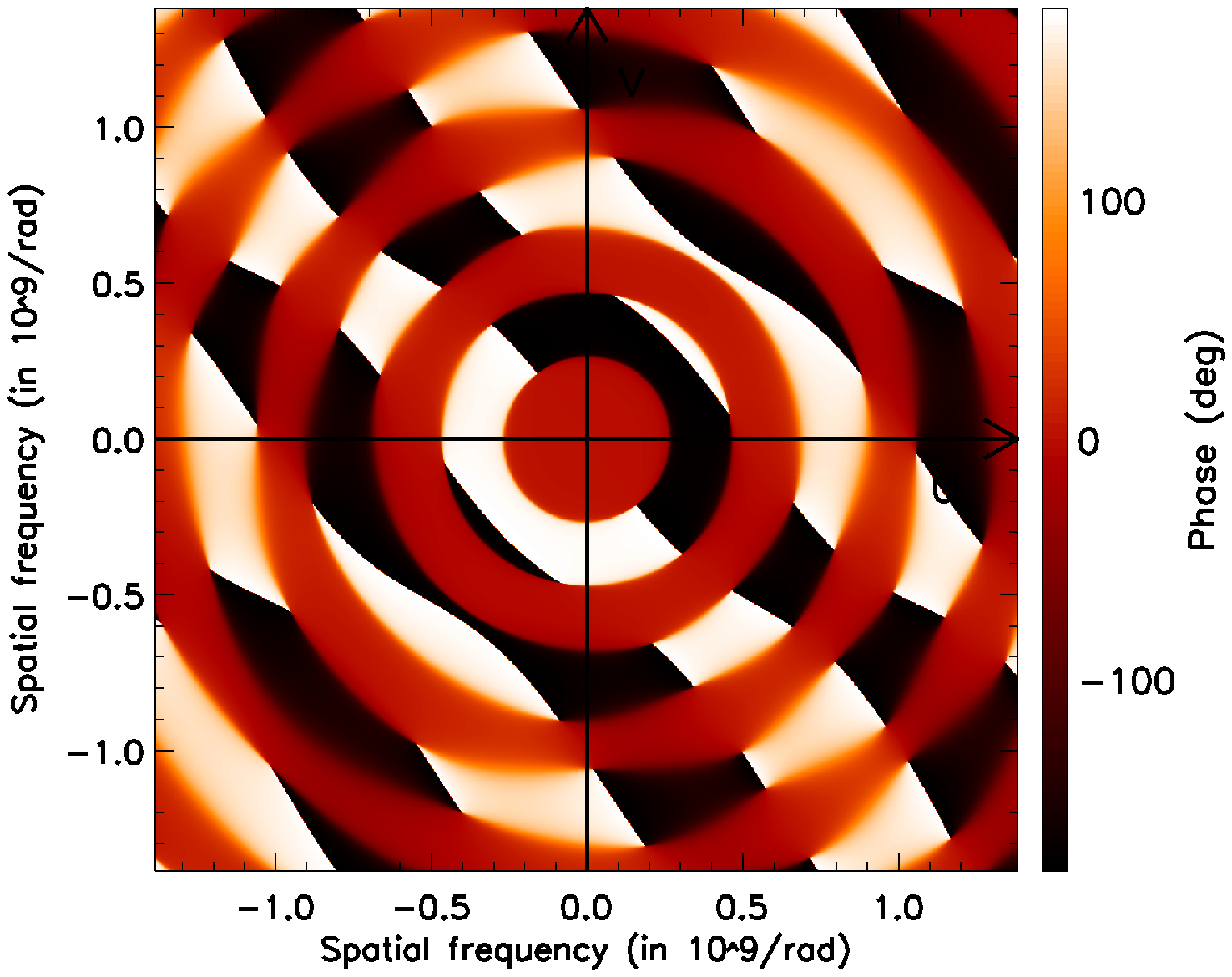} \\
\end{array}$
\end{center}
\caption{Map of the signal induced by an exoplanet at ($0.2, 0.0$)~mas and a spot at ($0.2, 0.2$)~mas in the visibility modulus (log scale, upper panels) and the phases (bottom panels), for a $1$~mas star. \textit{Left}: exoplanet alone with $\theta_{\rm p} = 0.10$~mas. \textit{Middle}: exoplanet of diameter $\theta_{\rm p} = 0.10$~mas with a spot of diameter $\theta_{\rm s} = 0.10$~mas. \textit{Right}: Same as middle with $\theta_{\rm p} = 0.05$~mas.}
\label{fig:Maps}
\end{figure*}

Figure~\ref{fig:Maps} (upper panels) shows the difference between the visibility of a single star and a star with an exoplanet and/or a spot, that is the absolute signal induced by these components. 

\begin{figure}
\centering
\includegraphics[scale=0.5]{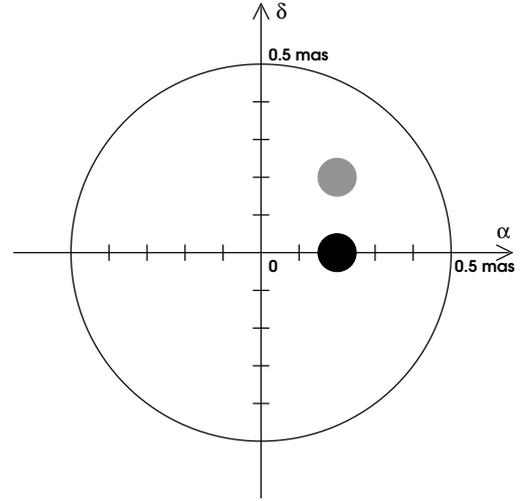}
\caption{Representation of the configuration of the system studied in Sec.4 for the case $\theta_{\rm p} = 0.10$~mas at ($0.2, 0.0$)~mas and $\theta_{\rm s} = 0.10$~mas at ($0.2, 0.2$)~mas. The black disk represents the exoplanet and the grey disk represents the spot.}
\label{fig:schema}
\end{figure}

The left panel represents the signal induced by an exoplanet alone, whose diameter and position are $\theta_{\rm p} = 0.10$~mas and ($0.2, 0.0$) (in mas). In the two other panels, we add a spot of diameter $\theta_{\rm s} = 0.10$~mas at the position ($0.2, 0.2$). In the middle panel, $\theta_{\rm p} = 0.10$~mas and in the right panel, $\theta_{\rm p} = 0.05$~mas. This is represented in Fig.~\ref{fig:schema}. If the accuracy of interferometers reaches the signal induced by both spot and exoplanet, their presence may affect the determination of stellar angular diameter because they are generally measured in this visibility lobe. But this would also mean that a simple limb-darkened representation of the star would not be accurate anymore \citep[see, e.g.,][]{Lacour2008}. At higher spatial frequencies, the interferometric signal becomes more sensitive to small structures (spots, granulation patterns) and three-dimensional models from radiative hydrodynamical atmospheric simulations become essential to fit the interferometric data \citep[see e.g.][]{Bigot2006, Lacour2008, Chiavassa2012, Chiavassa2014}. Using these models is another solution to take into account the stellar activity (in terms of stellar granulation) that also affects interferometric observables.

Figure~\ref{fig:Maps} (bottom panels) shows the corresponding phases. Again, the star with a spot and/or an exoplanet has a different signal from that of a single star, which varies according to the observing direction and sizes of the additional objects (the phase of a single star is $0 \pm 180^\circ$). When the spot and the exoplanet have the same size, the exoplanet signal is mixed up with the spot signal until the second lobe. When the exoplanet is smaller than the spot, the signal in the phase is totally disturbed by the spot, and their signal is already detectable from the second lobe. The exoplanet signal is thus hard to extract.

For both cases, not taking into account the spot leads to an incorrect interpretation of the signal, for example a larger transiting planet, and highlights the necessity of observing the star out of transit to characterize the stellar activity, which could be substracted from the signal recorded during the transit time to allow characterizing the exoplanet. Other effects should not be forgotten, such as the granulation, as shown by \cite{Chiavassa2014}, who uses a realistic three-dimensional radiative hydrodynamical simulations obtained with the stagger-grid \citep{Stagger2013}. It perturbs closure-phase measurements in particular from the third lobe of visibility. 
However, this would be possible to characterize a hot-Jupiter; characterizing an Earth-like planet could require a much higher sensitivity. 


\section{Discussion: exoplanet detection applied to CHARA and the VLTI}
\label{sec:discussion}

\begin{figure}[h]
\includegraphics[scale=0.5]{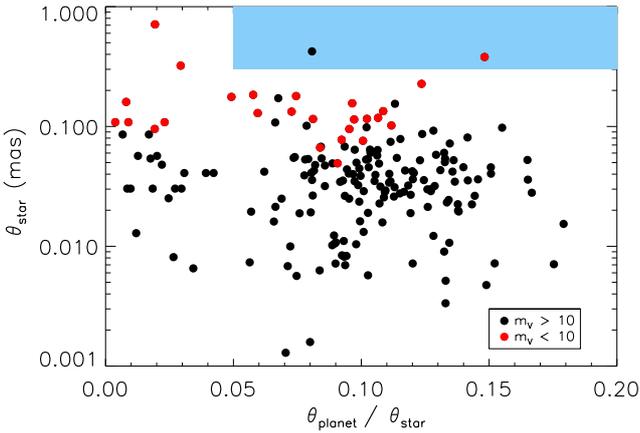}
\caption{Transiting exoplanet host stars with available distance and radius that allow deriving their angular diameter. They are plotted according to the ratio $\theta_{\rm p}/ \theta_{\star}$. Stars with $m_{\rm V} < 10$ are plotted in red and those with $m_{\rm V} > 10$ in black. The blue box represents the detecting ability of VEGA.}
\label{fig:ExoplanetsKnown}
\end{figure}

To evaluate the actual number of potential interferometric targets, and relying on the results of previous sections, we have to take into account the host star's angular diameter and the ratio $\theta_{\rm p}/\theta_{\star}$, as well as the stellar magnitude in the considered wavelength. Figure~\ref{fig:ExoplanetsKnown} shows all the transiting exoplanet host stars with available angular diameters according to the ratio $\theta_{\rm p}/\theta_{\star}$. Of the $\sim 400$ transiting exoplanets known today\footnote{January 2014}, only half of their parent stars have a known distance and/or known radius, which reduces the number of known angular diameter of transiting exoplanet host stars to $188$.

Most of transiting exoplanet host stars have a diameter of between $0.005$ and $0.20$~mas. The highest ratio $\theta_{\rm p}/\theta_{\star}$ is $\sim 0.15$, with sparse exoplanets (or stars) having larger angular diameters. If the signature of exoplanets as small as $\theta_{\rm p} = 0.05$~mas can be detected with baselines shorter than $200$~m under a high S/N (see Figs.~\ref{fig:MBLSqrdVisRel} and \ref{fig:Seuils}), the angular resolution already represents an important limitation in detecting of the stellar planetary companion. Since the visible domain provides a better angular resolution than the infrared domain at a similar baseline, interferometers operating in the visible wavelength appear to be more suitable. The CHARA array \citep{McAlister2012} holds the longest baselines in the world (up to $330$~m), and hosts many instruments, among them VEGA \citep{Mourard2009, Ligi2013} and PAVO \citep{Ireland2008}, which operate in the visible, and thus benefit from the best angular resolution in the world ($\sim 0.3$~mas), despite a limiting magnitude in the V band of around $\sim 8$. However, this resolution remains in principle too low as most of transiting exoplanet host stars have even smaller angular diameters. Phase-only measurements enable preliminary detections in some very specific cases. The NPOI is another good candidate to characterize exoplanets because it operates in the optical wavelength. Its baselines are currently too small to reach a high enough angular resolution (the longest baseline length is at present $79$~m), but it should be high enough within the next year \citep{Baines2014}.

In contrast, the VLTI does not own baselines longer than $\sim 130$~m, and since it mainly hosts instruments operating in the infrared domain, no instrument is able to reach such a resolution.
Furthermore, Fig. \ref{fig:ExoplanetsKnown} is affected by an instrumental bias toward low magnitudes because of detection instruments, that is the reason why the upper part is almost empty. Indeed, large surveys searching for exoplanets candidates were generally focused on faint stars (\textit{Kepler}, CoRoT), which are too faint to be observed by actual interferometers. Future missions should allow to fill this part of the diagram. SPHERE \citep{Beuzit2007}, a new instrument that is being developed for the VLT, is expected to detect giant planets around bright stars by direct imaging, but these planets will be rather far away from the stars, making the probability of a transit quite unlikely; SPHERE targets are also mostly young stars, with enhanced levels of activity. Future missions such as PLATO \citep[][$m_{V}<11$]{Rauer2012}, CHEOPS \citep[][$m_{V}<9$]{Broeg2013}, or TESS \citep[][$m_{V}<12$]{Ricker2010} will target bright stars, and complementary observations of these stars will become possible with interferometry and thus open new possibilities of exoplanet characterization with interferometry.

A few instruments are also being developed on CHARA and the VLTI. A new prototype of VEGA, called FRIEND (B\'erio et al., in prep), is currently being studied to reach a maximum angular resolution of $0.3$~mas and be able to observe stars up to $m_{\rm V} < 10$. They appear in red in Fig.~\ref{fig:ExoplanetsKnown}. Assuming these improvements, we see that only one known star becomes a potential target for this study (they are included in the blue box; the star represented by a black dot is too faint).
By only increasing the angular resolution of FRIEND by a factor of $3$, resulting in a maximum spatial resolution of $0.1$~mas, one can considerably increase the number of potential targets (by more than a factor of $10$). The baseline lengths and the sensitivity have to be improved together to reach the core of potential detectable exoplanets.

\section{Conclusion}

We have computed the interferometric observables of a star with the new code \texttt{COMETS} and observed the variations induced by a transiting exoplanet and/or a magnetic spot in squared visibilities and phases. We showed that several parameters of these objects affect the baseline length required to detect their signal, such as the position, diameter, or spot temperature. Starting from the variation of the MBL as a function of the exoplanet angular diameter, we derived an analytical solution that allows calculating the baseline length required to detect the absolute exoplanet signature in squared visibilities. This directly shows the interest of using the visible wavelength over the infrared, and that visible interferometers are better adapted for exoplanet characterization. More importantly, current instruments use long enough baselines to reach the spatial angular resolution that can lead to a characterization of large exoplanets or very dark spots, but lack accuracy on squared visibility and phase (or CP) measurements. 

To detect a $0.10$~mas exoplanet crossing a 1 mas star at visible wavelength, an accuracy better than $\sim 0.5\%$ from the first null is required in squared visibilities, at best observing conditions. A precision better than $\sim 1^{\circ}$ on phases is necessary in the first lobe, or better than $\sim 6^{\circ}$ in the second lobe. At present, no instrument can reach these accuracies. To detect a $0.05$~mas exoplanet, the accuracy needed on squared visibilities and phases are $\sim 0.1\%$ and $\sim 1^{\circ}$ from the first null. Magnetic spot signals can easily mimic these exoplanet signals, and distinguishing between them is essential for an exoplanet characterization, but requires measurements generally beyond the third lobe of visibility.

\begin{acknowledgements}
We thank the anonymous referee, whose constructive remarks led to a significant improvement of the paper. RL also acknowledges the Ph.D. financial support from the Observatoire de la C\^ote d'Azur and the PACA region and support from OCA after her Ph.D. This research has made use of The Extrasolar Planets Encyclopaedia at \texttt{exoplanet.eu}.
\end{acknowledgements}

\bibliographystyle{aa}
\bibliography{These}

\begin{thebibliography}{54}
\expandafter\ifx\csname natexlab\endcsname\relax\def\natexlab#1{#1}\fi

\bibitem[{{Absil} {et~al.}(2011){Absil}, {Le Bouquin}, {Berger}, {Lagrange},
  {Chauvin}, {Lazareff}, {Zins}, {Haguenauer}, {Jocou}, {Kern}, {Millan-Gabet},
  {Rochat}, \& {Traub}}]{Absil2011}
{Absil}, O., {Le Bouquin}, J.-B., {Berger}, J.-P., {et~al.} 2011, \aap, 535,
  A68

\bibitem[{{Absil} {et~al.}(2010){Absil}, {Le Bouquin}, {Lebreton}, {Augereau},
  {Benisty}, {Chauvin}, {Hanot}, {M{\'e}rand}, \& {Montagnier}}]{Absil2010}
{Absil}, O., {Le Bouquin}, J.-B., {Lebreton}, J., {et~al.} 2010, \aap, 520, L2

\bibitem[{{Armstrong} {et~al.}(1998){Armstrong}, {Mozurkewich}, {Rickard},
  {Hutter}, {Benson}, {Bowers}, {Elias}, {Hummel}, {Johnston}, {Buscher},
  {Clark}, {Ha}, {Ling}, {White}, \& {Simon}}]{NPOI}
{Armstrong}, J.~T., {Mozurkewich}, D., {Rickard}, L.~J., {et~al.} 1998, \apj,
  496, 550

\bibitem[{{Baines} {et~al.}(2014){Baines}, {Armstrong}, {Schmitt}, {Benson},
  {Zavala}, \& {van Belle}}]{Baines2014}
{Baines}, E.~K., {Armstrong}, J.~T., {Schmitt}, H.~R., {et~al.} 2014, \apj,
  781, 90

\bibitem[{{Baines} {et~al.}(2008){Baines}, {McAlister}, {ten Brummelaar},
  {Turner}, {Sturmann}, {Sturmann}, {Goldfinger}, \& {Ridgway}}]{Baines2008}
{Baines}, E.~K., {McAlister}, H.~A., {ten Brummelaar}, T.~A., {et~al.} 2008,
  \apj, 680, 728

\bibitem[{{Baines} {et~al.}(2007){Baines}, {van Belle}, {ten Brummelaar},
  {McAlister}, {Swain}, {Turner}, {Sturmann}, \& {Sturmann}}]{Baines2007}
{Baines}, E.~K., {van Belle}, G.~T., {ten Brummelaar}, T.~A., {et~al.} 2007,
  \apjl, 661, L195

\bibitem[{{Baron} {et~al.}(2012){Baron}, {Cotton}, {Lawson}, {Ridgway},
  {Aarnio}, {Monnier}, {Hofmann}, {Schertl}, {Weigelt}, {Thi{\'e}baut},
  {Soulez}, {Mary}, {Millour}, {Vannier}, {Young}, {Elias}, {Schmitt}, \&
  {Rengaswamy}}]{Baron2012}
{Baron}, F., {Cotton}, W.~D., {Lawson}, P.~R., {et~al.} 2012, in Society of
  Photo-Optical Instrumentation Engineers (SPIE) Conference Series, Vol. 8445

\bibitem[{{Berdyugina}(2005)}]{Berdyugina2005}
{Berdyugina}, S.~V. 2005, Living Reviews in Solar Physics, 2, 8

\bibitem[{{Beuzit} {et~al.}(2007){Beuzit}, {Feldt}, {Dohlen}, {Mouillet},
  {Puget}, {Antichi}, {Baudoz}, {Boccaletti}, {Carbillet}, {Charton}, {Claudi},
  {Fusco}, {Gratton}, {Henning}, {Hubin}, {Joos}, {Kasper}, {Langlois},
  {Moutou}, {Pragt}, {Rabou}, {Saisse}, {Schmid}, {Turatto}, {Udry}, {Vakili},
  {Waters}, \& {Wildi}}]{Beuzit2007}
{Beuzit}, J.-L., {Feldt}, M., {Dohlen}, K., {et~al.} 2007, in In the Spirit of
  Bernard Lyot: The Direct Detection of Planets and Circumstellar Disks in the
  21st Century

\bibitem[{{Bigot} {et~al.}(2006){Bigot}, {Kervella}, {Th{\'e}venin}, \&
  {S{\'e}gransan}}]{Bigot2006}
{Bigot}, L., {Kervella}, P., {Th{\'e}venin}, F., \& {S{\'e}gransan}, D. 2006,
  \aap, 446, 635

\bibitem[{{Broeg} {et~al.}(2013){Broeg}, {Fortier}, {Ehrenreich}, {Alibert},
  {Baumjohann}, {Benz}, {Deleuil}, {Gillon}, {Ivanov}, {Liseau}, {Meyer},
  {Oloffson}, {Pagano}, {Piotto}, {Pollacco}, {Queloz}, {Ragazzoni}, {Renotte},
  {Steller}, \& {Thomas}}]{Broeg2013}
{Broeg}, C., {Fortier}, A., {Ehrenreich}, D., {et~al.} 2013, in European
  Physical Journal Web of Conferences, Vol.~47, European Physical Journal Web
  of Conferences, 3005

\bibitem[{{Che} {et~al.}(2012){Che}, {Monnier}, {Kraus}, {Baron}, {Pedretti},
  {Thureau}, \& {Webster}}]{Che2012}
{Che}, X., {Monnier}, J.~D., {Kraus}, S., {et~al.} 2012, in Society of
  Photo-Optical Instrumentation Engineers (SPIE) Conference Series, Vol. 8445

\bibitem[{{Chelli} {et~al.}(2009){Chelli}, {Duvert}, {Malbet}, \&
  {Kern}}]{Chelli2009}
{Chelli}, A., {Duvert}, G., {Malbet}, F., \& {Kern}, P. 2009, \aap, 498, 321

\bibitem[{{Chiavassa} {et~al.}(2012){Chiavassa}, {Bigot}, {Kervella}, {Matter},
  {Lopez}, {Collet}, {Magic}, \& {Asplund}}]{Chiavassa2012}
{Chiavassa}, A., {Bigot}, L., {Kervella}, P., {et~al.} 2012, \aap, 540, A5

\bibitem[{{Chiavassa} {et~al.}(2014){Chiavassa}, {Ligi}, {Magic}, {Collet},
  {Asplund}, \& {Mourard}}]{Chiavassa2014}
{Chiavassa}, A., {Ligi}, R., {Magic}, Z., {et~al.} 2014, \aap, 567, A115

\bibitem[{{Claret} \& {Bloemen}(2011)}]{Claret2011}
{Claret}, A. \& {Bloemen}, S. 2011, \aap, 529, A75

\bibitem[{{Coud{\'e} du Foresto} {et~al.}(2003){Coud{\'e} du Foresto}, {Borde},
  {Merand}, {Baudouin}, {Remond}, {Perrin}, {Ridgway}, {ten Brummelaar}, \&
  {McAlister}}]{Coude2003}
{Coud{\'e} du Foresto}, V., {Borde}, P.~J., {Merand}, A., {et~al.} 2003, in
  Society of Photo-Optical Instrumentation Engineers (SPIE) Conference Series,
  ed. W.~A. {Traub}, Vol. 4838, 280--285

\bibitem[{{Duvert} {et~al.}(2010){Duvert}, {Chelli}, {Malbet}, \&
  {Kern}}]{Duvert2010}
{Duvert}, G., {Chelli}, A., {Malbet}, F., \& {Kern}, P. 2010, \aap, 509, A66

\bibitem[{{Garcia} {et~al.}(2014){Garcia}, {van Belle}, {Muterspaugh}, \&
  {Swihart}}]{Garcia2014}
{Garcia}, E., {van Belle}, G., {Muterspaugh}, M.~W., \& {Swihart}, S. 2014, in
  American Astronomical Society Meeting Abstracts, Vol. 223, American
  Astronomical Society Meeting Abstracts \#223, 154.26

\bibitem[{{Howell} {et~al.}(2012){Howell}, {Rowe}, {Bryson}, {Quinn}, {Marcy},
  {Isaacson}, {Ciardi}, {Chaplin}, {Metcalfe}, {Monteiro}, {Appourchaux},
  {Basu}, {Creevey}, {Gilliland}, {Quirion}, {Stello}, {Kjeldsen},
  {Christensen-Dalsgaard}, {Elsworth}, {Garc{\'{\i}}a}, {Houdek}, {Karoff},
  {Molenda-{\.Z}akowicz}, {Thompson}, {Verner}, {Torres}, {Fressin}, {Crepp},
  {Adams}, {Dupree}, {Sasselov}, {Dressing}, {Borucki}, {Koch}, {Lissauer},
  {Latham}, {Buchhave}, {Gautier}, {Everett}, {Horch}, {Batalha}, {Dunham},
  {Szkody}, {Silva}, {Mighell}, {Holberg}, {Ballot}, {Bedding}, {Bruntt},
  {Campante}, {Handberg}, {Hekker}, {Huber}, {Mathur}, {Mosser}, {R{\'e}gulo},
  {White}, {Christiansen}, {Middour}, {Haas}, {Hall}, {Jenkins}, {McCaulif},
  {Fanelli}, {Kulesa}, {McCarthy}, \& {Henze}}]{Howell2012}
{Howell}, S.~B., {Rowe}, J.~F., {Bryson}, S.~T., {et~al.} 2012, \apj, 746, 123

\bibitem[{{Huber} {et~al.}(2012){Huber}, {Ireland}, {Bedding}, {Howell},
  {Maestro}, {M{\'e}rand}, {Tuthill}, {White}, {Farrington}, {Goldfinger},
  {McAlister}, {Schaefer}, {Sturmann}, {Sturmann}, {ten Brummelaar}, \&
  {Turner}}]{Huber2012}
{Huber}, D., {Ireland}, M.~J., {Bedding}, T.~R., {et~al.} 2012, \mnras, 423,
  L16

\bibitem[{{Ireland} {et~al.}(2008){Ireland}, {M{\'e}rand}, {ten Brummelaar},
  {Tuthill}, {Schaefer}, {Turner}, {Sturmann}, {Sturmann}, \&
  {McAlister}}]{Ireland2008}
{Ireland}, M.~J., {M{\'e}rand}, A., {ten Brummelaar}, T.~A., {et~al.} 2008, in
  Society of Photo-Optical Instrumentation Engineers (SPIE) Conference Series,
  Vol. 7013

\bibitem[{{Labeyrie} {et~al.}(2012{\natexlab{a}}){Labeyrie}, {Allouche},
  {Mourard}, {Bolgar}, {Chakraborty}, {Maillot}, {Palitzyne}, {Poletti},
  {Rochaix}, {Prud'homme}, {Rondi}, {Roussel}, \& {Surya}}]{Labeyrie2012a}
{Labeyrie}, A., {Allouche}, F., {Mourard}, D., {et~al.} 2012{\natexlab{a}}, in
  Society of Photo-Optical Instrumentation Engineers (SPIE) Conference Series,
  Vol. 8445

\bibitem[{{Labeyrie} {et~al.}(2012{\natexlab{b}}){Labeyrie}, {Mourard},
  {Allouche}, {Chakraborthy}, {Dejonghe}, {Surya}, {Bresson}, {Aime}, {Mary},
  \& {Carlotti}}]{Labeyrie2012b}
{Labeyrie}, A., {Mourard}, D., {Allouche}, F., {et~al.} 2012{\natexlab{b}}, in
  Society of Photo-Optical Instrumentation Engineers (SPIE) Conference Series,
  Vol. 8445

\bibitem[{{Lacour} {et~al.}(2008){Lacour}, {Meimon}, {Thi{\'e}baut}, {Perrin},
  {Verhoelst}, {Pedretti}, {Schuller}, {Mugnier}, {Monnier}, {Berger},
  {Haubois}, {Poncelet}, {Le Besnerais}, {Eriksson}, {Millan-Gabet}, {Ragland},
  {Lacasse}, \& {Traub}}]{Lacour2008}
{Lacour}, S., {Meimon}, S., {Thi{\'e}baut}, E., {et~al.} 2008, \aap, 485, 561

\bibitem[{{Le Bouquin} {et~al.}(2011){Le Bouquin}, {Berger}, {Lazareff},
  {Zins}, {Haguenauer}, {Jocou}, {Kern}, {Millan-Gabet}, {Traub}, {Absil},
  {Augereau}, {Benisty}, {Blind}, {Bonfils}, {Bourget}, {Delboulbe},
  {Feautrier}, {Germain}, {Gitton}, {Gillier}, {Kiekebusch}, {Kluska},
  {Knudstrup}, {Labeye}, {Lizon}, {Monin}, {Magnard}, {Malbet}, {Maurel},
  {M{\'e}nard}, {Micallef}, {Michaud}, {Montagnier}, {Morel}, {Moulin},
  {Perraut}, {Popovic}, {Rabou}, {Rochat}, {Rojas}, {Roussel}, {Roux},
  {Stadler}, {Stefl}, {Tatulli}, \& {Ventura}}]{LeBouquin2011}
{Le Bouquin}, J.-B., {Berger}, J.-P., {Lazareff}, B., {et~al.} 2011, \aap, 535,
  A67

\bibitem[{{Leinert} {et~al.}(2003){Leinert}, {Graser}, {Przygodda}, {Waters},
  {Perrin}, {Jaffe}, {Lopez}, {Bakker}, {B{\"o}hm}, {Chesneau}, {Cotton},
  {Damstra}, {de Jong}, {Glazenborg-Kluttig}, {Grimm}, {Hanenburg}, {Laun},
  {Lenzen}, {Ligori}, {Mathar}, {Meisner}, {Morel}, {Morr}, {Neumann}, {Pel},
  {Schuller}, {Rohloff}, {Stecklum}, {Storz}, {von der L{\"u}he}, \&
  {Wagner}}]{Leinert2003}
{Leinert}, C., {Graser}, U., {Przygodda}, F., {et~al.} 2003, \apss, 286, 73

\bibitem[{{Lhom{\'e}} {et~al.}(2012){Lhom{\'e}}, {Scott}, {ten Brummelaar},
  {Mollier}, {Reess}, {Chapron}, {Buey}, {Sevin}, {Sturmann}, {Sturmann}, \&
  {Coud{\'e} du Foresto}}]{Lhome2012}
{Lhom{\'e}}, E., {Scott}, N., {ten Brummelaar}, T., {et~al.} 2012, in Society
  of Photo-Optical Instrumentation Engineers (SPIE) Conference Series, Vol.
  8445

\bibitem[{{Ligi} {et~al.}(2012){Ligi}, {Mourard}, {Lagrange}, {Perraut},
  {Boyajian}, {B{\'e}rio}, {Nardetto}, {Tallon-Bosc}, {McAlister}, {ten
  Brummelaar}, {Ridgway}, {Sturmann}, {Sturmann}, {Turner}, {Farrington}, \&
  {Goldfinger}}]{Ligi2012}
{Ligi}, R., {Mourard}, D., {Lagrange}, A.~M., {et~al.} 2012, \aap, 545, A5

\bibitem[{Ligi {et~al.}(2013)Ligi, Mourard, Nardetto, \& Clausse}]{Ligi2013}
Ligi, R., Mourard, D., Nardetto, N., \& Clausse, J.-M. 2013, Journal of
  Astronomical Instrumentation, 02, 1340003

\bibitem[{{Lopez}(2012)}]{Lopez2012}
{Lopez}, B. 2012, Publications de l'Observatoire Astronomique de Beograd, 91,
  129

\bibitem[{{Maestro} {et~al.}(2012){Maestro}, {Kok}, {Huber}, {Ireland},
  {Tuthill}, {White}, {Schaefer}, {ten Brummelaar}, {McAlister}, {Turner},
  {Farrington}, \& {Goldfinger}}]{Maestro2012}
{Maestro}, V., {Kok}, Y., {Huber}, D., {et~al.} 2012, in Society of
  Photo-Optical Instrumentation Engineers (SPIE) Conference Series, Vol. 8445

\bibitem[{{Magic} {et~al.}(2013){Magic}, {Collet}, {Asplund}, {Trampedach},
  {Hayek}, {Chiavassa}, {Stein}, \& {Nordlund}}]{Stagger2013}
{Magic}, Z., {Collet}, R., {Asplund}, M., {et~al.} 2013, \aap, 557, A26

\bibitem[{{Matter} {et~al.}(2010){Matter}, {Vannier}, {Morel}, {Lopez},
  {Jaffe}, {Lagarde}, {Petrov}, \& {Leinert}}]{Matter2010}
{Matter}, A., {Vannier}, M., {Morel}, S., {et~al.} 2010, \aap, 515, A69

\bibitem[{{Mayor} \& {Queloz}(1995)}]{Mayor1995}
{Mayor}, M. \& {Queloz}, D. 1995, \nat, 378, 355

\bibitem[{{McAlister} {et~al.}(2012){McAlister}, {ten Brummelaar}, {Ridgway},
  {Gies}, {Sturmann}, {Sturmann}, {Turner}, {Schaefer}, {Boyajian},
  {Farrington}, {Goldfinger}, \& {Webster}}]{McAlister2012}
{McAlister}, H.~A., {ten Brummelaar}, T.~A., {Ridgway}, S.~T., {et~al.} 2012,
  in Society of Photo-Optical Instrumentation Engineers (SPIE) Conference
  Series, Vol. 8445

\bibitem[{{Monnier} {et~al.}(2004){Monnier}, {Berger}, {Millan-Gabet}, \& {ten
  Brummelaar}}]{Monnier2004}
{Monnier}, J.~D., {Berger}, J.-P., {Millan-Gabet}, R., \& {ten Brummelaar},
  T.~A. 2004, in Society of Photo-Optical Instrumentation Engineers (SPIE)
  Conference Series, ed. W.~A. {Traub}, Vol. 5491, 1370

\bibitem[{{Mourard} {et~al.}(2009){Mourard}, {Clausse}, {Marcotto}, {Perraut},
  {Tallon-Bosc}, {B{\'e}rio}, {Blazit}, {Bonneau}, {Bosio}, {Bresson},
  {Chesneau}, {Delaa}, {H{\'e}nault}, {Hughes}, {Lagarde}, {Merlin}, {Roussel},
  {Spang}, {Stee}, {Tallon}, {Antonelli}, {Foy}, {Kervella}, {Petrov},
  {Thiebaut}, {Vakili}, {McAlister}, {ten Brummelaar}, {Sturmann}, {Sturmann},
  {Turner}, {Farrington}, \& {Goldfinger}}]{Mourard2009}
{Mourard}, D., {Clausse}, J.~M., {Marcotto}, A., {et~al.} 2009, \aap, 508, 1073

\bibitem[{{Nutzman} {et~al.}(2011){Nutzman}, {Fabrycky}, \&
  {Fortney}}]{Nutzman2011}
{Nutzman}, P.~A., {Fabrycky}, D.~C., \& {Fortney}, J.~J. 2011, \apjl, 740, L10

\bibitem[{{Petrov} \& {AMBER Consortium}(2003)}]{Petrov2003}
{Petrov}, R.~G. \& {AMBER Consortium}. 2003, in EAS Publications Series,
  Vol.~6, EAS Publications Series, ed. G.~{Perrin} \& F.~{Malbet}, 111

\bibitem[{{Rauer} \& {Catala}(2012)}]{Rauer2012}
{Rauer}, H. \& {Catala}, C. 2012, in EGU General Assembly Conference Abstracts,
  Vol.~14, EGU General Assembly Conference Abstracts, ed. A.~{Abbasi} \&
  N.~{Giesen}, 7033

\bibitem[{{Ricker} {et~al.}(2010){Ricker}, {Latham}, {Vanderspek}, {Ennico},
  {Bakos}, {Brown}, {Burgasser}, {Charbonneau}, {Clampin}, {Deming}, {Doty},
  {Dunham}, {Elliot}, {Holman}, {Ida}, {Jenkins}, {Jernigan}, {Kawai},
  {Laughlin}, {Lissauer}, {Martel}, {Sasselov}, {Schingler}, {Seager},
  {Torres}, {Udry}, {Villasenor}, {Winn}, \& {Worden}}]{Ricker2010}
{Ricker}, G.~R., {Latham}, D.~W., {Vanderspek}, R.~K., {et~al.} 2010, in
  Bulletin of the American Astronomical Society, Vol.~42, American Astronomical
  Society Meeting Abstracts $\#$215, 450.06

\bibitem[{{Sanchis-Ojeda} {et~al.}(2011){Sanchis-Ojeda}, {Winn}, {Holman},
  {Carter}, {Osip}, \& {Fuentes}}]{Sanchis2011}
{Sanchis-Ojeda}, R., {Winn}, J.~N., {Holman}, M.~J., {et~al.} 2011, \apj, 733,
  127

\bibitem[{Scott {et~al.}(2013)Scott, Millan-Gabet, Lhom\'é, Brrummelaar,
  Foresto, Sturmann, \& Sturmann}]{Scott2013}
Scott, N.~J., Millan-Gabet, R., Lhom\'é, E., {et~al.} 2013, Journal of
  Astronomical Instrumentation, 02, 1340005

\bibitem[{{Silva-Valio} \& {Lanza}(2011)}]{Silva2011}
{Silva-Valio}, A. \& {Lanza}, A.~F. 2011, \aap, 529, A36

\bibitem[{{Strassmeier}(2009)}]{Strassmeier2009}
{Strassmeier}, K.~G. 2009, \aapr, 17, 251

\bibitem[{{ten Brummelaar} {et~al.}(2012){ten Brummelaar}, {Sturmann},
  {McAlister}, {Sturmann}, {Turner}, {Farrington}, {Schaefer}, {Goldfinger}, \&
  {Kloppenborg}}]{tenBrummelaar2012}
{ten Brummelaar}, T.~A., {Sturmann}, J., {McAlister}, H.~A., {et~al.} 2012, in
  Society of Photo-Optical Instrumentation Engineers (SPIE) Conference Series,
  Vol. 8445

\bibitem[{{Torres} {et~al.}(2008){Torres}, {Winn}, \& {Holman}}]{Torres2008}
{Torres}, G., {Winn}, J.~N., \& {Holman}, M.~J. 2008, \apj, 677, 1324

\bibitem[{{Traub} \& {Jucks}(2002)}]{Traub2002}
{Traub}, W.~A. \& {Jucks}, K.~W. 2002, Washington DC American Geophysical Union
  Geophysical Monograph Series, 130, 369

\bibitem[{{van Belle}(2008)}]{vanBelle2008}
{van Belle}, G.~T. 2008, \pasp, 120, 617

\bibitem[{{van Leeuwen}(2007)}]{vanLeeuwen2007}
{van Leeuwen}, F. 2007, \aap, 474, 653

\bibitem[{{Zhao} {et~al.}(2011){Zhao}, {Monnier}, {Che}, {Pedretti}, {Thureau},
  {Schaefer}, {ten Brummelaar}, {M{\'e}rand}, {Ridgway}, {McAlister}, {Turner},
  {Sturmann}, {Sturmann}, {Goldfinger}, \& {Farrington}}]{Zhao2011}
{Zhao}, M., {Monnier}, J.~D., {Che}, X., {et~al.} 2011, \pasp, 123, 964

\bibitem[{{Zhao} {et~al.}(2010){Zhao}, {Monnier}, {Che}, {ten Brummelaar},
  {Pedretti}, \& {Thureau}}]{Zhao2010}
{Zhao}, M., {Monnier}, J.~D., {Che}, X., {et~al.} 2010, in Society of
  Photo-Optical Instrumentation Engineers (SPIE) Conference Series, Vol. 7734

\bibitem[{{Zhao} {et~al.}(2008){Zhao}, {Monnier}, {ten Brummelaar}, {Pedretti},
  \& {Thureau}}]{Zhao2008}
{Zhao}, M., {Monnier}, J.~D., {ten Brummelaar}, T., {Pedretti}, E., \&
  {Thureau}, N. 2008, in IAU Symposium, Vol. 249, IAU Symposium, ed. Y.-S.
  {Sun}, S.~{Ferraz-Mello}, \& J.-L. {Zhou}, 71--77

\end{thebibliography}

\Online
\begin{appendix}
\section{Calculating the complex visibility}

\subsection{Transiting planet}
\label{annexPlanet}
We have calculated the $2$D Fourier transform (TF) of the intensity profile $\tilde{I}$. The normalized $\tilde{I}$ gives the complex visibility. The resulting TF of the stellar profile $\tilde{I}_{\star} (\vec{\rho})$ and of the exoplanet profile $\tilde{I}_{\rm p} (\vec{\rho})$ are
\begin{equation}
 \begin{aligned}
  \label{equ:DSstar}
	\tilde{I}_{\star} (\vec{\rho}) =  I_{\star}(1) \frac{\pi \theta_{\star}^2}{4} \left(1- \frac{b}{3} \right) \frac{ \left[ a J_1(z)/z + b(\pi /2)^{1/2} J_{3/2}(z) / z^{3/2} \right]}	{(a /2 + b/3)} \\
 \end{aligned}
\end{equation}
\begin{equation}
\begin{aligned}
	\tilde{I}_{\rm p} (\vec{\rho}) &= \frac{\theta_{\star}}{2}  \frac{J_1 \left(\pi \theta_{\rm p} \rho\right)}{\rho} \\
	& \times \underbrace{\cos(2 \pi (u \alpha_{\rm p} + v \delta_{\rm p})) + i \sin (2 \pi (u \alpha_{\rm p} + v \delta_{\rm p}))}_N \ ,
 \end{aligned}
\end{equation}
with $a=1-b$, $z=\pi \theta_{\star} \rho$ and $\rho = B_{\rm p}/ \lambda$, $B_{\rm p}$ being the projected baseline and $\lambda$ the observing wavelength in nm. The spatial frequencies are represented by $u=\frac{B_x}{\lambda}$ and $v=\frac{B_y}{\lambda}$.
The resulting complex visibility of the system is
\begin{equation}
 \begin{aligned}
	V_{\rm p}(\rho) &= \left[ \tilde{I}_{\star}(\rho) - \left( I_{\star}(\mu_{\rm p}) - I_{\rm p} \right) \tilde{I}_{\rm p}(\rho) \right] / \\
	&\left[ \frac{\pi}{4} \left( I_{\star}(1) \theta_{\star}^2 \left( 1- \frac{b}{3} \right) - \left( I_{\star}(\mu_{\rm p}) - I_{\rm p} \right) \theta_{\rm p}^2 \right) \right] \ .
 \end{aligned}
\end{equation}

\subsection{Spot}
\label{annexSpot}
The normalized $2$D Fourier transform of Eq.~\ref{eq:IntensiteTache} is the complex visibility of the system $star + spot$.
The TF of the penumbra and umbra profiles are $\tilde{I}_{\rm pen}$ and $\tilde{I}_{\rm um}$
\begin{equation}
	\begin{aligned}
	\tilde{I}_{\rm pen}(\vec{\rho}) &= \frac{\theta_{\rm pen}} {2} \frac{J_{1}(\pi \theta_{\rm pen} \rho)} {\rho} \times N' \\
	\tilde{I}_{\rm um}(\vec{\rho}) &= \frac{\theta_{\rm um}}{2} \frac{J_{1}(\pi \theta_{\rm um} \rho)} {\rho} \times N' \ ,
	\end{aligned}
	\label{equ:DSOmbrePenombre}
\end{equation}
with $N' = \cos(2 \pi (u \alpha_{\rm s} + v \delta_{\rm s})) + i \sin (2 \pi (u \alpha_{\rm s} + v \delta_{\rm s}))$.
The TF of the stellar profile is written in Eq.~\ref{equ:DSstar}.
Combining these three TFs gives the final complex visibility:
\begin{equation}
	\begin{aligned}
 	V_{\rm s}(\rho) &= \left[ \tilde{I}_{\star}(\rho) - \left( I_{\star}(\mu_{\rm s}) - I_{\rm pen} \right) \tilde{I}_{\rm pen}(\vec{\rho}) \right. \\
 			 &\left. + \left( I_{\rm um} - I_{\rm pen} \right) \tilde{I}_{\rm um}(\vec{\rho}) \right] / \\
 			 & \left[ \frac{\pi}{4} \left( I_{\star}(1) \theta_{\star}^2 \left( 1- \frac{b}{3} \right) \right. \right. \\
 			 &\left. \left. - \left( I_{\star}(\mu_{\rm s}) - I_{\rm pen} \right) \theta_{\rm pen}^2 + \left( I_{\rm um} - I_{\rm pen} \right) \theta_{\rm um}^2 \right) \right] \ .
	\end{aligned}
\end{equation}

\subsection{Transiting planet and a spot}
\label{annexPlanetSpot}
The normalized TF of Eq.\ref{equ:intensityProfileTotal} gives the complex visibility of the system.
The TF of the exoplanet and the spot profiles are calculated in Sects.~\ref{subsec:exoplanet} and \ref{subsec:Spot}:
\begin{equation}
	\begin{aligned}
	\tilde{I}_{\rm pen}(\vec{\rho}) &= \frac{\theta_{\rm pen}} {2} \frac{J_{1}(\pi \theta_{\rm pen} \rho)} {\rho} \times N' \\
	\tilde{I}_{\rm um}(\vec{\rho}) &= \frac{\theta_{\rm um}}{2} \frac{J_{1}(\pi \theta_{\rm um} \rho)} {\rho} \times N' \\
	\tilde{I}_{\rm p} (\vec{\rho}) &= \frac{\theta_{\rm p}}{2}  \frac{J_1 \left(\pi \theta_{\rm p} \rho\right)}{\rho}   \times N \ .
	\end{aligned}
	\label{equ:DSExopl+Tache}
\end{equation}		
The final complex visibility is thus
\begin{equation}
	\begin{aligned}
 	V_{\rm p+s}(\rho) &= \left[ \tilde{I}_{\star}(\vec{\rho}) -  \left( I_{\star}(\mu_{\rm p}) - I_{\rm p} \right) \tilde{I}_{\rm p} (\vec{\rho}) \right. \\
 	& \left. -(I_{\star}(\mu_{\rm s}) - I_{\rm pen}) \tilde{I}_{\rm pen}(\vec{\rho}) \right. \\
 	& \left. + (I_{\rm um} - I_{\rm pen}) \tilde{I}_{\rm um}(\vec{\rho}) \right] /  \\
 				  &   \left[ \frac{\pi}{4} \left( I_{\star}(1) \theta_{\star}^2 \left( 1- \frac{b}{3}\right) - ( I_{\star}(\mu_{\rm p}) - I_{\rm p}) \theta_{\rm p}^2 \right. \right.\\
 				  & \left. \left. - (I_{\star}(\mu_{\rm s}) - I_{\rm pen}) \theta_{\rm pen}^2 + (I_{\rm um} - I_{\rm pen}) \theta_{\rm um}^2  \right) \right] \ ,
	\end{aligned}
\end{equation}
for which we use the squared modulus.

\end{appendix}

\end{document}